\documentclass[aps,prd,nofootinbib,preprint]{revtex4}

\usepackage{amssymb}
\usepackage{amsmath,color}
\usepackage{epsfig}
\usepackage{graphicx,epsf,epsfig}
\usepackage{bbm}
\usepackage{subfigure}
\usepackage{multirow}
\usepackage{setspace}

\newcommand\T{\rule{0pt}{2.8ex}} 
\newcommand\B{\rule[-1.2ex]{0pt}{0pt}} 
\newcommand{\ul}{\underline}
\newcommand{\mee}{\langle m_{ee}\rangle}
\newcommand{\sumnu}{\sum m_i}
\newcommand{\mbeta}{m_\beta}
\newcommand{\dms}{\Delta m_{\rm S}^2}
\newcommand{\dma}{\Delta m_{\rm A}^2}
\newcommand{\dmnewa}{\Delta m_{41}^2}
\newcommand{\dmnewb}{\Delta m_{51}^2}
\newcommand{\sssol}{\sin^2\!\theta_{12}}
\newcommand{\ssatm}{\sin^2\!\theta_{23}}
\newcommand{\ssre}{\sin^2\!\theta_{13}}
\newcommand{\sssta}{\sin^2\!\theta_{14}}
\newcommand{\ssstb}{\sin^2\!\theta_{24}}
\newcommand{\ssstc}{\sin^2\!\theta_{34}}

\newcommand{\obb}{0\nu\beta\beta}
\newcommand{\ba}{\begin{array}{c}}
\newcommand{\baz}{\begin{array}{cc}}
\newcommand{\bad}{\begin{array}{ccc}}
\newcommand{\bav}{\begin{array}{cccc}}
\newcommand{\baf}{\begin{array}{ccccc}}
\newcommand{\ea}{\end{array}}

\def\be{\begin{equation}}
\def\ee{\end{equation}}
\def\gs{\mathrel{
   \rlap{\raise 0.511ex \hbox{$>$}}{\lower 0.511ex \hbox{$\sim$}}}}
\def\ls{\mathrel{
   \rlap{\raise 0.511ex \hbox{$<$}}{\lower 0.511ex \hbox{$\sim$}}}}

\newcommand{\bea}{\begin{equation} \begin{array}{c}}
\newcommand{\eea}{ \end{array} \end{equation}}
\newcommand{\D}{\displaystyle}

\def\slc#1{\setbox0=\hbox{$#1$}           
    \dimen0=\wd0                                 
    \setbox1=\hbox{/} \dimen1=\wd1               
    \ifdim\dimen0>\dimen1                        
       \rlap{\hbox to \dimen0{\hfil/\hfil}}      
       #1                                        
    \else                                        
       \rlap{\hbox to \dimen1{\hfil$#1$\hfil}}   
       /                                         
    \fi}
\begin{document}
\mbox{}\vspace{2cm}
\title{\Large Light Sterile Neutrinos: Models and Phenomenology}
\mbox{}\vspace{1cm}

\author{James Barry}
\email{james.barry@mpi-hd.mpg.de}

 \affiliation{Max-Planck-Institut f{\"u}r Kernphysik, Postfach
 103980, 69029 Heidelberg, Germany}

\author{Werner Rodejohann}
\email{werner.rodejohann@mpi-hd.mpg.de}

 \affiliation{Max-Planck-Institut f{\"u}r Kernphysik, Postfach
 103980, 69029 Heidelberg, Germany}

\author{He Zhang}
\email{he.zhang@mpi-hd.mpg.de}

\affiliation{Max-Planck-Institut f{\"u}r Kernphysik, Postfach
103980, 69029 Heidelberg, Germany}


\begin{abstract}
\noindent 
Motivated by recent hints in particle physics and cosmology, we study the realization of eV-scale sterile neutrinos within both the seesaw mechanism and flavor symmetry theories. We show that light sterile neutrinos can rather easily be accommodated in the popular $A_4$ flavor symmetry models. The exact tri-bimaximal mixing pattern is perturbed due to active-sterile mixing, which we discuss in detail for one example. In addition, we find an interesting extension of the type I seesaw, which can provide a natural origin for eV-scale sterile neutrinos as well as visible admixtures between sterile and active neutrinos. We also show that the presence of sterile neutrinos would significantly change the observables in neutrino experiments, specifically the oscillation probabilities in short-baseline experiments and the effective mass in neutrino-less double beta decay. The latter can prove particularly helpful in strengthening the case for eV-scale sterile neutrinos. 

\end{abstract}
\maketitle

\section{Introduction}

Neutrino oscillation experiments have provided firm evidence that, in contrast to the prediction of the Standard Model
(SM), neutrinos are massive and that their flavors change during propagation. Their unusual mixing pattern and the smallness of their masses makes the explanation of the origin of neutrino masses and leptonic flavor mixing one of the most challenging problems in particle physics. In the standard neutrino oscillation picture three active neutrinos are involved, with mass-squared differences of order $10^{-4}$ and $10^{-3}~{\rm eV}^2$. The absolute mass scale of neutrinos is also
constrained to be less than around $1~{\rm eV}$ from tritium beta decay experiments as well as cosmological observations. For a recent review on our current understanding of neutrino observables, see Ref.~\cite{GonzalezGarcia:2007ib}. 

Despite the successful achievements of solar, atmospheric, reactor and accelerator neutrino experiments, there are experimental anomalies that cannot be explained within the standard three neutrino framework. In particular, the possible presence of sterile neutrinos points towards non-standard neutrino physics. The issue of the LSND and MiniBooNE results has been around for some time, and is frequently interpreted as a hint towards the presence of one or two sterile neutrino states \cite{Sorel:2003hf,Maltoni:2007zf,Karagiorgi:2009nb,Giunti:2010wz}. Recently this debate has been re-ignited by a reevaluation of the anti-neutrino spectra of nuclear reactors \cite{Mention:2011rk}, which leads to increased fluxes. As a result, the negative results of previous reactor experiments can in fact be interpreted as the observation of a flux deficit, and this in turn can be explained by additional sterile neutrinos with masses at the eV scale \cite{Mention:2011rk,Kopp:2011qd}. Interestingly, current results from precision cosmology and Big Bang nucleosynthesis mildly favor extra radiation in the Universe beyond photons and ordinary neutrinos. While this could be any relativistic degree of freedom, the interpretation in terms of additional sterile neutrino species is straightforward. Indeed, several cosmological parameter fits (e.g., the analyses of the CMB or SDSS data sets in Ref.~\cite{Hamann:2010bk}) have been found to be compatible with more radiation than that predicted by the SM particle content. 
This is supported by the recently reported higher $^4$He abundance \cite{Izotov:2010ca,Aver:2010wq}, which in the framework of Big Bang nucleosynthesis can be accommodated by additional relativistic degrees of freedom, as it leads to earlier freeze-out of the weak reactions, resulting in a higher neutron-to-proton ratio. It is rather intriguing that hints for the presence of sterile neutrinos are given by fundamentally different probes: neutrino oscillations, Big Bang nucleosynthesis and observation of the Universe's structure. This is the situation which motivates the present study. 

Sterile neutrinos, if they exist, would lead to rich experimental phenomena. By definition, sterile neutrinos do not
directly enter the weak interactions. However, their admixture with active neutrinos would modify the neutrino flavor mixing and lead to observable effects in neutrino oscillation experiments. Furthermore, due to that admixture they could interact with gauge bosons, resulting in significant corrections to certain non-oscillation processes, e.g., in the neutrino-less double decay ($\obb$) amplitude \cite{Bilenky:2001xq,Goswami:2005ng,Goswami:2007kv} or in beta decay spectra, such as in the KATRIN experiment \cite{Riis:2010zm,Barrett:2011jg}. It has also been pointed out that an eV-scale sterile neutrino would significantly affect the the atmospheric neutrino fluxes in the energy range 500 GeV to a few TeV, and these effects could be studied in the IceCube detector~\cite{Nunokawa:2003ep,Choubey:2007ji,Razzaque:2011ab}. Other aspects that are in principle modified by such sterile states are supernova physics \cite{Choubey:2007ga}, solar neutrinos \cite{Palazzo:2011rj}, or the interpretation of cosmological data \cite{Kristiansen:2011mp}. We will add to this discussion by updating the predictions for $\obb$ with the recent results for the relevant mixing parameters. We consider all possible neutrino mass spectra, namely four in case of one sterile neutrino, and eight in case of two sterile neutrinos. For the latter case we also point out some properties of short-baseline oscillation probabilities in four often overlooked possible spectra.

On the other hand, we consider the non-trivial theoretical origin of eV-scale sterile neutrinos. In what regards neutrino masses and their peculiar mixing structure, the key words are seesaw mechanism \cite{Minkowski:1977sc,Yanagida:1979as,GellMann:1980vs,Mohapatra:1979ia} and flavor symmetries (see \cite{Altarelli:2010gt,Ishimori:2010au} for recent reviews). We will show that popular models based on the $A_4$
flavor symmetry can easily be modified to take sterile neutrinos into account. As a result, the mixing structure of the original model, tri-bimaximal mixing (TBM) in our example, is perturbed, and corrections to the tri-bimaximal values arise. This will be a general feature of such approaches, but here we focus on one concrete example, which we will outline in detail.\footnote{Note that a general analysis of the mixing of active and sterile neutrinos has been presented in Ref.~\cite{Smirnov:2006bu}.}

Sterile neutrinos are a necessary ingredient of the canonical type I seesaw mechanism, though they are naturally assumed to be many orders of magnitude heavier than the SM scale of $10^2$ GeV. As a result, their mixing with the SM particles is highly suppressed. If one brings one of the heavy states down to the eV scale, it is possible to generate a sterile neutrino with the correct mass and (potentially) the correct mixing with the SM leptons to explain the data. However, another more non-trivial case can also be studied, namely an extension of the type I seesaw with additional heavy (i.e.~heavier than the SM scale) neutral fermions. This approach generates sterile neutrinos without the need to initially have states with eV masses, and is more in the seesaw spirit.\\  

The remaining parts of this work are organized as follows: in Sec.~\ref{sec:phenomena}, we outline the formalism of neutrino mixing in the presence of sterile states and summarize the phenomenological consequences of sterile neutrinos in neutrino-less double beta decay. Short comments on short-baseline neutrino oscillations are delegated to the Appendix. Then, in Sec.~\ref{sec:A4}, we discuss how to embed sterile neutrinos into one particular $A_4$ flavor symmetry model, and describe the resulting deviations from exact TBM. Section~\ref{sec:model} is devoted to a general overview of the realization of sterile neutrinos in seesaw models. Finally, we conclude in Sec.~\ref{sec:summary}. The individual sections are largely independent of each other, but their results and methods could be combined. However, we feel that a separate discussion of each aspect is more suitable for the present discourse.

\section{Phenomenological consequences of sterile neutrinos}
\label{sec:phenomena}
In this section we will outline some properties of neutrino parameters and evaluate the contributions to the effective mass relevant for neutrino-less double beta decay. 

\subsection{Neutrino mixing with sterile neutrinos} \label{sec:mixing}

In the presence of $n_s = n-3$ sterile neutrinos, the neutrino mass matrix
is an $n\times n$ matrix $m_\nu$, which can be diagonalized by means
of an $n\times n$ unitary matrix $U$. The neutrino flavor eigenstates 
$\nu_f$ (for $f=e,\mu,\tau,s_1, s_2,\ldots,s_{n-3}$) 
are then related to their mass eigenstates $\nu_i$ (for
$i=1,2,\ldots,n$) via 
\begin{eqnarray}
\nu_f = \sum^n_{i=1} U_{fi} \nu_i \, .
\end{eqnarray}
In general, for $n$ massive families including  $n_s = n-3 \neq 0$ massive sterile neutrinos,
one has $n - 1 = n_s + 2$ Majorana phases, $3 (n-2) = 3 (n_s + 1)$ mixing angles and 
$2 n - 5 = 2 n_s + 1$ Dirac phases. The number of angles and Dirac
phases is less than the naive $\frac 12 n (n-1)$ angles and 
$\frac12 (n-1) (n-2)$ phases, because the 
$\frac 12 n_s (n_s - 1)$ rotations between sterile states are unphysical. 
For illustration, in the case of only one sterile neutrino, $U$ is
typically parameterized by 
\begin{equation}
U = R_{34}\tilde{R}_{24}\tilde{R}_{14}R_{23}\tilde{R}_{13}R_{12}P \,
, \label{eq:UPMNS4x4}
\end{equation}
where the matrices $R_{ij}$ are rotations in $ij$ space, i.e.
\begin{equation}
R_{34} = \begin{pmatrix} 1 & 0 & 0 & 0 \\ 0 & 1 & 0 & 0 \\ 0 & 0 &
c_{34} & s_{34} \\ 0 & 0 & -s_{34} & c_{34} \end{pmatrix} \quad {\rm
or} \quad \tilde{R}_{14} = \begin{pmatrix} c_{14} & 0 & 0 &
s_{14}e^{-i\delta_{14}} \\ 0 & 1 & 0 & 0 \\ 0 & 0 & 1 & 0 \\
-s_{14}e^{i\delta_{14}} & 0 & 0 & c_{14} \end{pmatrix} ,
\end{equation}
where $s_{ij} = \sin\theta_{ij}$, $c_{ij}=\cos\theta_{ij}$. 
The diagonal $P$ matrix contains the three Majorana phases $\alpha$,
$\beta$ and $\gamma$:
\begin{equation} P = {\rm
diag}\left(1,e^{i\alpha/2},e^{i(\beta/2+\delta_{13})},e^{i(\gamma/2+\delta_{14})}\right)\, .
\end{equation}
Note that there are in total three Dirac CP-violating phases
$\delta_{ij}$. The above definition of $P$ is constructed in such a way that only Majorana phases show up
in the effective mass governing neutrino-less double beta decay (see below). Similarly, one can parameterize the mixing matrix for 2 sterile neutrinos as 
\be \label{eq:U2st}
U = \tilde{R}_{25} R_{34} R_{25} \tilde{R}_{24} R_{23} \tilde{R}_{15}
\tilde{R}_{14} \tilde{R}_{13} R_{12} P\, ,
\ee
where $P = {\rm diag}(1,e^{i\alpha/2},e^{i(\beta/2 +
\delta_{13})},e^{i(\gamma/2 + \delta_{14})},
e^{i(\phi/2 + \delta_{15})})$.\\  

The mass-squared differences associated with the current hints for sterile neutrinos are much larger than the ones responsible for solar and atmospheric oscillations. The sterile neutrinos are thus separated in mass from the active ones and can either be heavier or lighter than the active ones. The active neutrinos can be normally ($m_3 > m_2 > m_1$) or inversely ($m_2 > m_1 > m_3$) ordered. Therefore, if there is one sterile neutrino, there are in total four possible mass orderings. Our nomenclature for the schemes is such that we go from top to bottom, i.e.~if $m_s \gg m_{1,2,3}$, we denote this 1+3 scenario as ``SN'' if the active neutrinos are normally ordered and ``SI'' if they are inversely ordered. Analogously, if the sterile state is lighter than the active ones ($m_{1,2,3} \gg m_s$, 3+1 scenario), we name the schemes ``NS'' and ``IS'', respectively. Obviously, 3+1 scenarios are less attractive since they induce (more) tension with the cosmological bound on the sum of neutrino masses. The models we will present in later sections indeed predict 1+3 scenarios. 

In the case of two sterile neutrinos with masses $m_{s_1}$ and $m_{s_2}$, there are three classes of mass spectra: $m_{s_1}, m_{s_2} \gg m_{1,2,3}$ (2+3 scenarios), $m_{1,2,3} \gg m_{s_1},m_{s_2}$ (3+2) and \cite{Goswami:2007kv} $m_{s_2} \gg m_{1,2,3} \gg m_{s_1}$ (1+3+1). The latter class of 1+3+1 orderings, first noted in Ref.~\cite{Goswami:2007kv}, has received 
less attention in the past, and was for instance not taken into account in the fits of Refs.~\cite{Sorel:2003hf,Maltoni:2007zf,Karagiorgi:2009nb}. The 1+3+1 cases have the interesting property that the active states are sandwiched between the two sterile ones. Current oscillation data cannot distinguish 1+3 from 3+1, or
3+2 from 2+3 scenarios, but are sensitive to 1+3+1 vs.~2+3/3+2 (see the Appendix). Ref.~\cite{Kopp:2011qd} fits the world's short-baseline data also within the 1+3+1 scenario and finds that the fit is slightly better than that for 2+3/3+2. The 2+3 cases are denoted ``SSN'' or ``SSI'', while the 3+2 cases are called ``NSS'' or ``ISS''. In what regards the 1+3+1 cases, two possible permutations of the mass spectrum should be distinguished. One of the sterile neutrinos is separated from the active ones by a larger mass gap, associated with the larger of the two mass-squared differences. If the heavier of the two sterile neutrinos is connected with the larger mass-squared difference, the scenarios are denoted ''SNSa'' and ``SISa'',  respectively; if the heavier sterile state is connected with the smaller mass-squared difference, we call the cases ''SNSb'' and ``SISb''. The 2+3 scenarios are more attractive than the others because they predict a smaller sum of masses. The individual masses $m_{1,2,3,4,5}$ expressed in terms of the mass-squared differences $\dms$, $\dma$,  $\dmnewa$ and $\dmnewb$ can be found in Ref.~\cite{Goswami:2007kv}, where the generalization to three sterile neutrinos (with sixteen possible mass orderings) is also discussed. 

\begin{table}[t]
 \centering
 \caption{Best-fit (from Ref.~\cite{Kopp:2011qd}) and estimated $2\sigma$ values of the sterile neutrino parameters.}
 \label{table:osc_params}
 \vspace{2mm}
 \begin{tabular}{lccccc}
  \hline \hline & parameter & $\dmnewa$ [eV] & $|U_{e4}|^2$ & $\dmnewb$ [eV] & $|U_{e5}|^2$ \\ \hline
 \multirow{2}{*}{3+1/1+3} & best-fit & 1.78 & 0.023 & & \\
 & $2\sigma$ & 1.61--2.01 & 0.006--0.040 & & \\ \hline
 \multirow{2}{*}{3+2/2+3} & best-fit & 0.47 & 0.016 & 0.87 & 0.019  \\
  & $2\sigma$ & 0.42--0.52 & 0.004--0.029 & 0.77--0.97 & 0.005--0.033
\\ \hline
\multirow{2}{*}{1+3+1} & best-fit & 0.47 & 0.017 & 0.87 & 0.020  \\
  & $2\sigma$ & 0.42--0.52 & 0.004--0.029 & 0.77--0.97 & 0.005--0.035 \\
\hline \hline
 \end{tabular}
\end{table}

Table~\ref{table:osc_params} shows the best-fit and $2\sigma$ ranges of the relevant parameters used for the analysis in this work. The best-fit values are taken from the global fit in Table II of Ref.~\cite{Kopp:2011qd}. In their analysis of the 3+1/1+3 scenarios, the authors of Ref.~\cite{Kopp:2011qd} find several different allowed regions in the $\dmnewa-\sssta$ parameter space, at $2\sigma$. For convenience we use the region around the best-fit point, and (since the ranges are not available) allow the parameters in the 3+2, 2+3 and 1+3+1 cases to have $2\sigma$ uncertainties of the same relative magnitude as those in the 3+1/1+3 scenario. The data favor the presence of two sterile neutrinos, mostly because they allow different neutrino and anti-neutrino probabilities, thus alleviating the tension between the LSND and MiniBooNE results. As mentioned above, 1+3+1 scenarios have a slightly better fit than 3+2/2+3 cases. 

We note in addition that the recent results from the 
T2K \cite{Abe:2011sj} and 
MINOS \cite{MINOSnew} experiments strengthen the existing hints \cite{Fogli:2008jx,Maltoni:2008ka,Schwetz:2011qt} for non-zero $\theta_{13}$, and that the analysis in Ref.~\cite{Fogli:2011qn} finds evidence for $\theta_{13} > 0$ at the level of $> 3\sigma$. It is not yet evident whether these new data will improve or worsen the fits in the various sterile neutrino scenarios; on the other hand it can also be argued that the T2K result is not due to $\theta_{13}$ but is actually another signature of sterile neutrinos \cite{Gibin:2011hd}.

\subsection{Neutrino-less double beta decay}

Neutrino-less double beta decay ($\obb$) is the only realistic test of lepton number violation. While there are several mechanisms to mediate the process (e.g.~heavy neutrinos, right-handed currents or SUSY particles), light Majorana neutrino exchange is presumably the best motivated scenario. We will work in this standard interpretation of $\obb$, and study the effect of massive sterile neutrinos in the eV range. This section is an update of the results in Refs.~\cite{Goswami:2005ng,Goswami:2007kv}; we also use the best-fit values in Table~\ref{table:osc_params} to report approximate numerical lower limits for the complementary mass observables $m_\beta\equiv\sqrt{|U_{ei}|^2m_i^2}$ and $\sumnu$, constrained by beta decay and cosmology, respectively.

\begin{figure}[t]
\centering
\subfigure{\label{fig:mee1p3}
\includegraphics[width=0.8\textwidth]{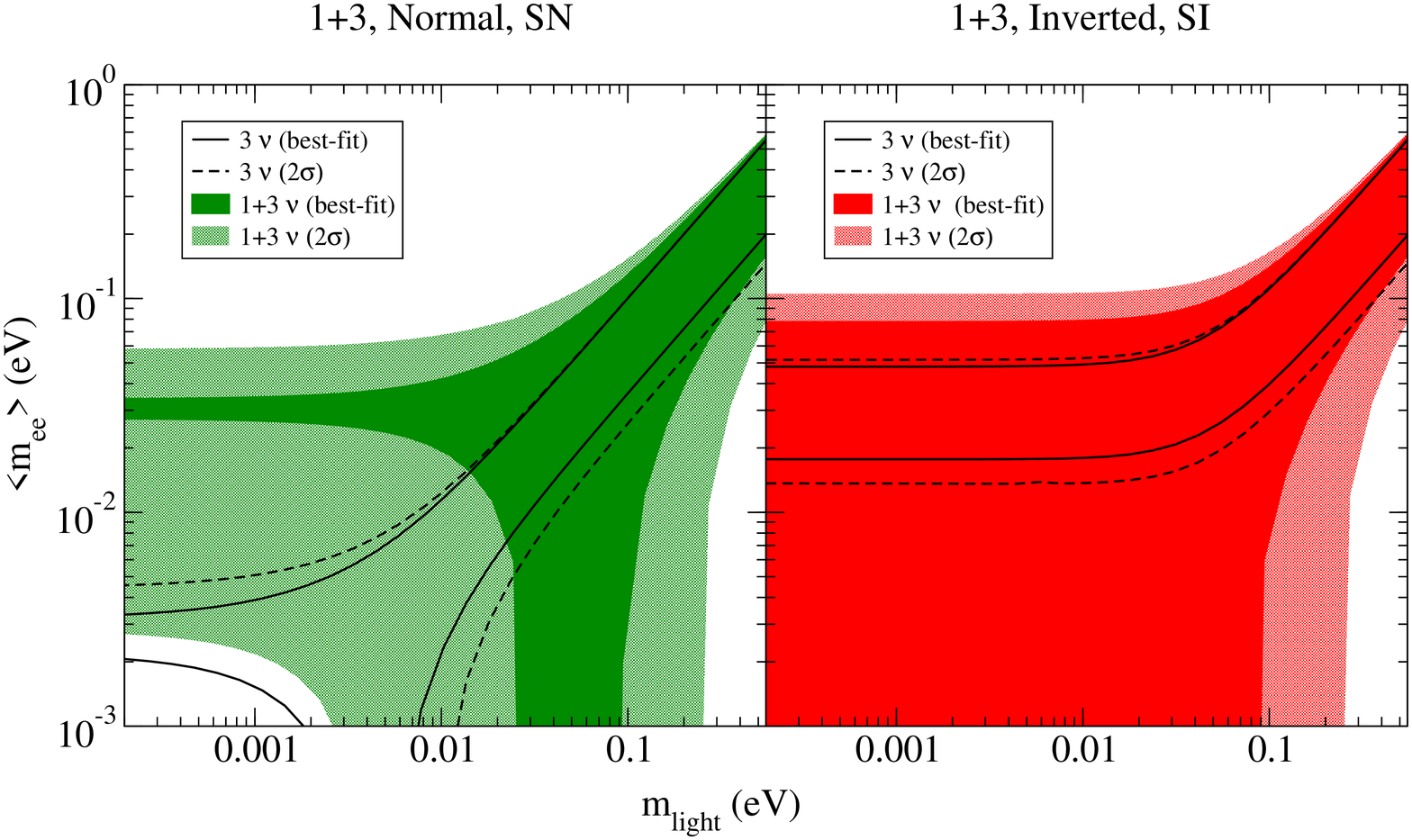}}
\subfigure{\label{fig:mee3p1}
\includegraphics[width=0.8\textwidth]{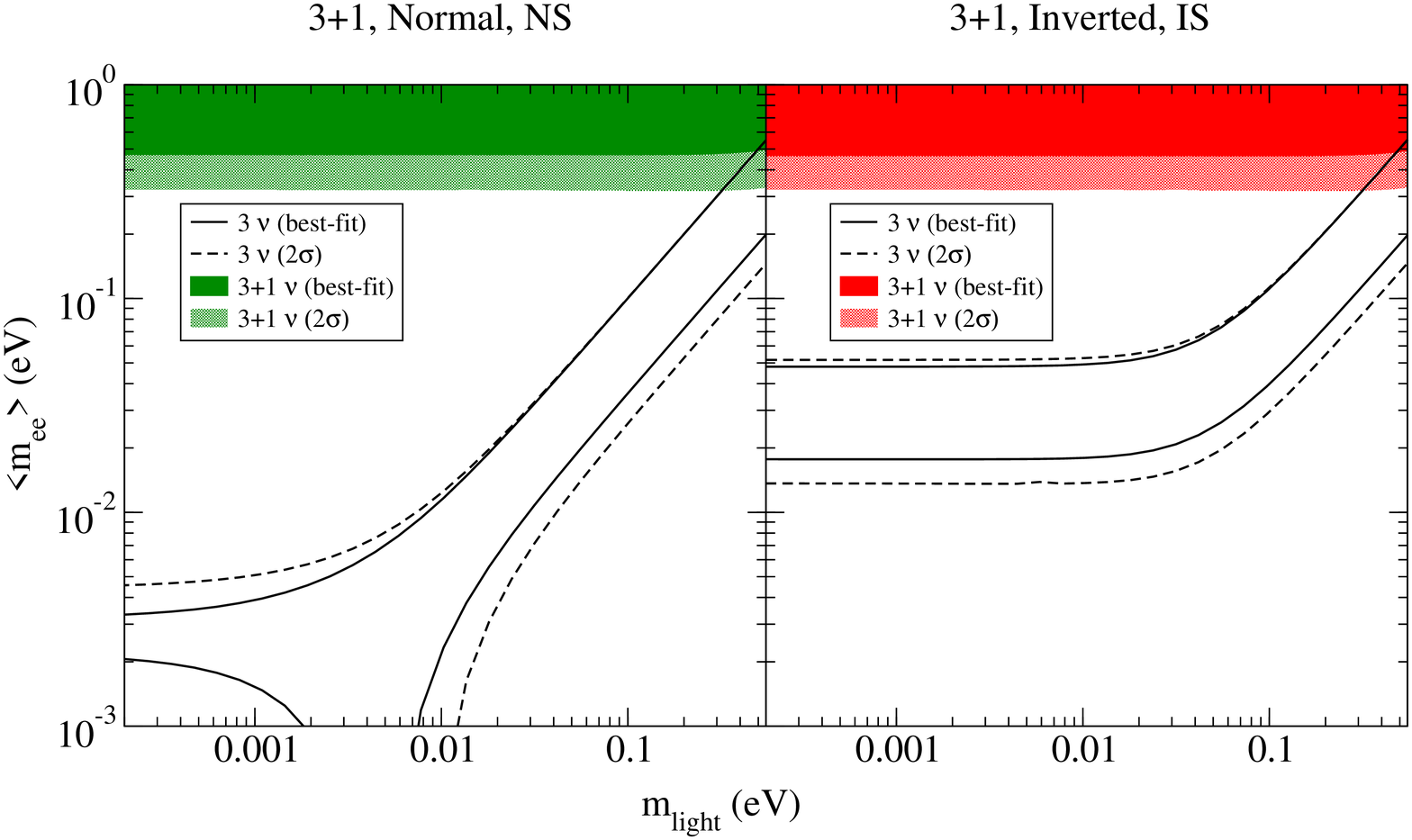}}
\caption{\label{fig:mee_4nu}The allowed ranges in the $\mee-m_{\rm
light}$ parameter space, both in the standard three-neutrino picture
(unshaded regions) and with one sterile neutrino (shaded regions), for
the 1+3 (top) and 3+1 (bottom) cases.}
\end{figure}

In the presence of one sterile neutrino, the effective neutrino mass in $\obb$ is given by
\begin{equation}\label{eq:meff4}
\mee_{4\nu} =
\left|c_{12}^2c_{13}^2c_{14}^2m_1+s_{12}^2c_{13}^2c_{14}^2m_2e^{i\alpha}
+s_{13}^2c_{14}^2m_3e^{i\beta}+s_{14}^2m_4e^{i\gamma}\right|  ,
\end{equation}
using the parameterization in Eq.~\eqref{eq:UPMNS4x4}. 
If the sterile neutrino is heavier than the active ones, the approximation
\begin{equation}
 \mee_{(1+3)\nu} \simeq \left| c_{14}^2 \mee_{3\nu} + s_{14}^2\sqrt{\dmnewa}e^{i\gamma}\right|
\label{eq:mee_3p1}
\end{equation}
holds, where $\mee_{3\nu} =
c_{12}^2c_{13}^2m_1+s_{12}^2c_{13}^2m_2e^{i\alpha}+s_{13}^2m_3e^{i\beta}$
is the standard expression for three active neutrinos. 
The upper panel of Fig.~\ref{fig:mee_4nu} displays the allowed range of $\mee_{(1+3)\nu}$ as a
function of the lightest mass $m_{\rm light}$, using data from
Refs.~\cite{Schwetz:2011qt,Kopp:2011qd}. Also shown in this plot and
the following ones is the allowed range of $|\mee_{3\nu}|$, i.e.~the
standard case. Its crucial features (see \cite{Rodejohann:2011mu} and references therein) are that $|\mee_{3\nu}|$ can vanish exactly in the normal hierarchy case, but cannot vanish in neither the inverted hierarchy nor the quasi-degenerate case. This standard behavior can be completely mixed up by the presence of one or more sterile neutrinos: if the lightest
neutrino mass is reasonably small, e.g.~$m_{\rm light} < 0.01~{\rm eV}$, the allowed range of $\mee_{(1+3)\nu}$ is dominated by the term $s_{14}^2\sqrt{\dmnewa} \simeq 0.031~{\rm eV}$, which means that $\mee_{(1+3)\nu}$ cannot vanish in the
normal ordering case (the contribution of the two light active neutrinos cannot cancel that of the sterile neutrino). However, in the inverted ordering $\mee_{(1+3)\nu}$ can vanish even for very small active neutrino masses. The effective mass can also be zero in the regime where the active neutrinos are quasi-degenerate ($m_{\rm light} > 0.1~{\rm eV}$). 

This modified behavior of $\mee$ is of particular interest in the
inverted ordering case, since the usual lower bound on the effective
mass  (cf. the solid and dashed lines in Fig.~\ref{fig:mee_4nu}) is
no longer valid. If future $\obb$ experiments measure a tiny effective
mass and the neutrino mass hierarchy is confirmed to be inverted from
long-baseline neutrino oscillations, the sterile neutrino hypothesis
would be an attractive explanation for this inconsistency. In the 3+1 case, $m_\beta$ and $\sumnu$ are dominated by the sterile contribution, i.e. $m_\beta \gs \sqrt{|U_{e4}|^2\dmnewa} \simeq 0.2$~eV and $\sumnu \gs \sqrt{\dmnewa} \simeq 1.3$ eV, respectively.

\begin{figure}[htp]
\centering
\subfigure{\label{fig:mee2p3}
\includegraphics[width=0.8\textwidth]{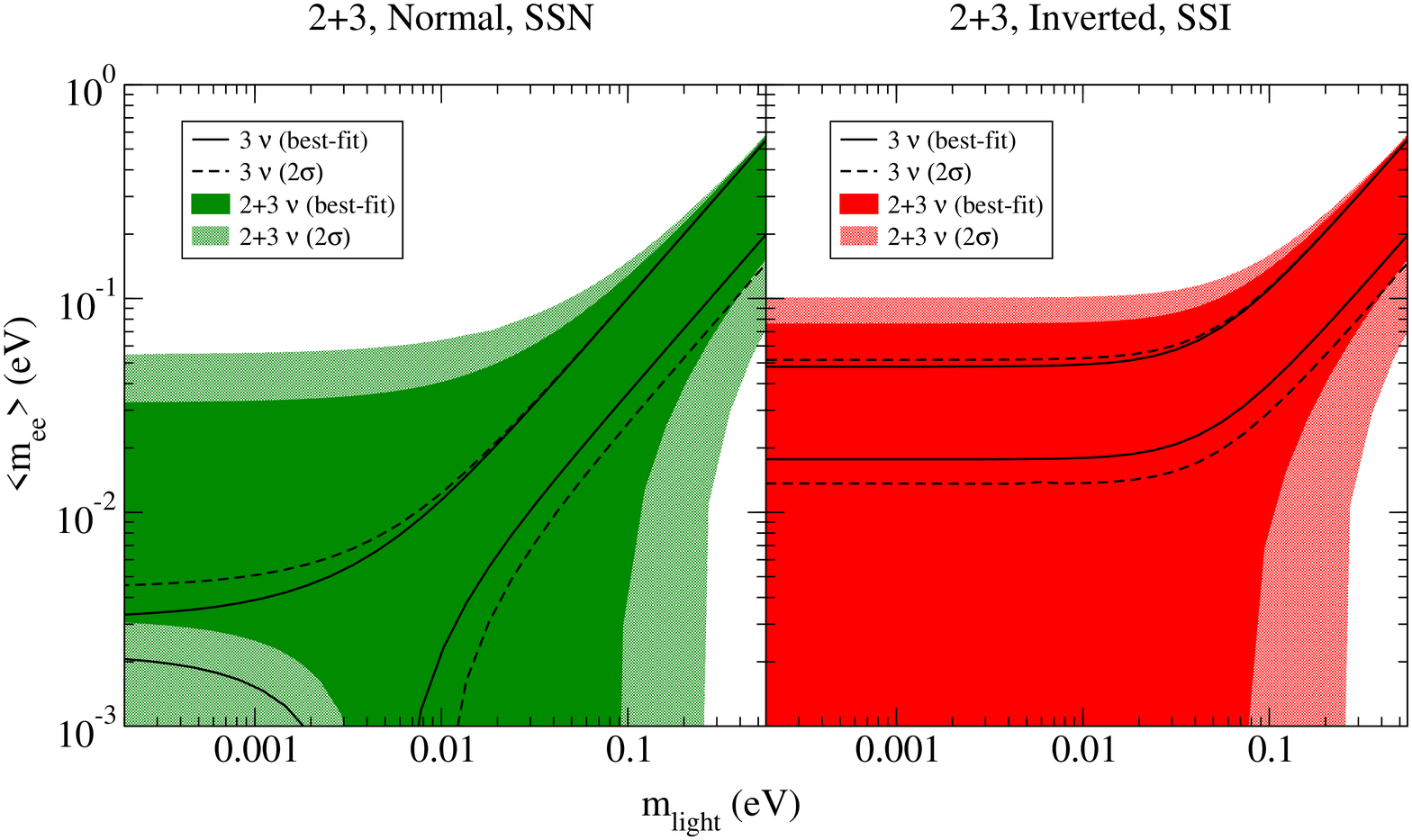}}
\subfigure{\label{fig:mee3p2}
\includegraphics[width=0.8\textwidth]{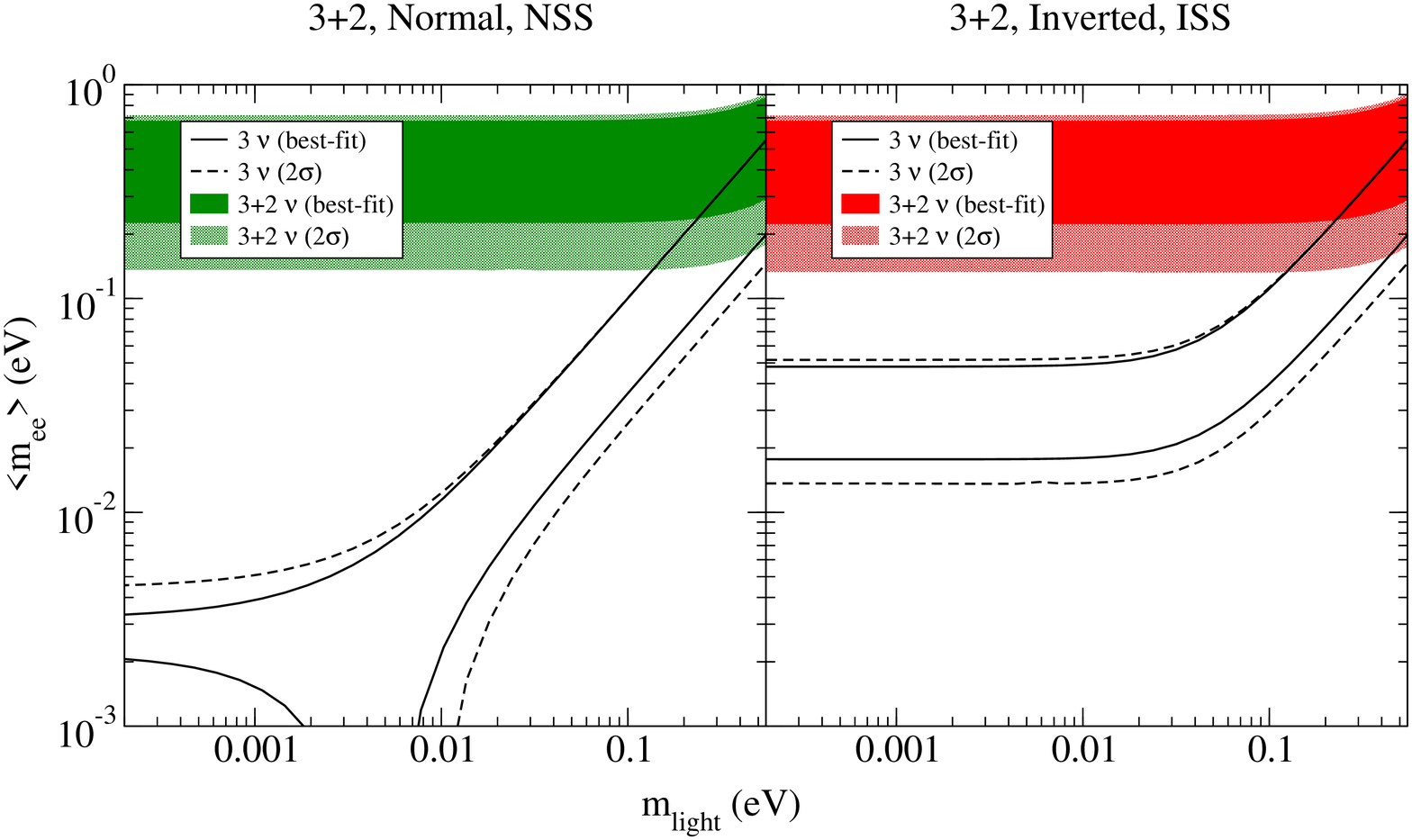}}
\caption{\label{fig:mee_5nu_a}Same as Fig.~\ref{fig:mee_4nu}, for the
2+3 (top) and 3+2 (bottom) cases.}
\end{figure}

The lower panel of Fig.~\ref{fig:mee_4nu} shows the effective mass
when the sterile neutrino is lighter than the active ones (the 3+1
scenario). In this case there are three quasi-degenerate neutrinos at
the eV scale, with their mass given by $\sqrt{\dmnewa} \simeq 1.3$ eV (this mass scale governs predictions for $\mbeta$). The effective mass then takes its standard form for quasi-degenerate neutrinos: 
\begin{equation} \label{eq:meffQD}
\mee_{(3+1)\nu} \simeq \sqrt{\dmnewa} \, \sqrt{1 - \sin^2 2
\theta_{12} \, \sin^2 \alpha/2} 
\, . 
\end{equation}
However, this situation is relatively disfavored by cosmological
bounds on the sum of neutrino masses \cite{Hamann:2010bk}, since $\sumnu \gs 3\sqrt{\dmnewa} \simeq 4$ eV. If taken at
face value, the current limit of about 0.5 eV on the effective mass means that $\sqrt{1 -\sin^2 2 \theta_{12} \, \sin^2 \alpha/2} \ls  0.4 $, thus already putting strong constraints on the solar neutrino mixing angle and in particular the Majorana phase.\\ 

\begin{figure}[tp]
\centering
\subfigure{\label{fig:mee131a}
\includegraphics[width=0.8\textwidth]{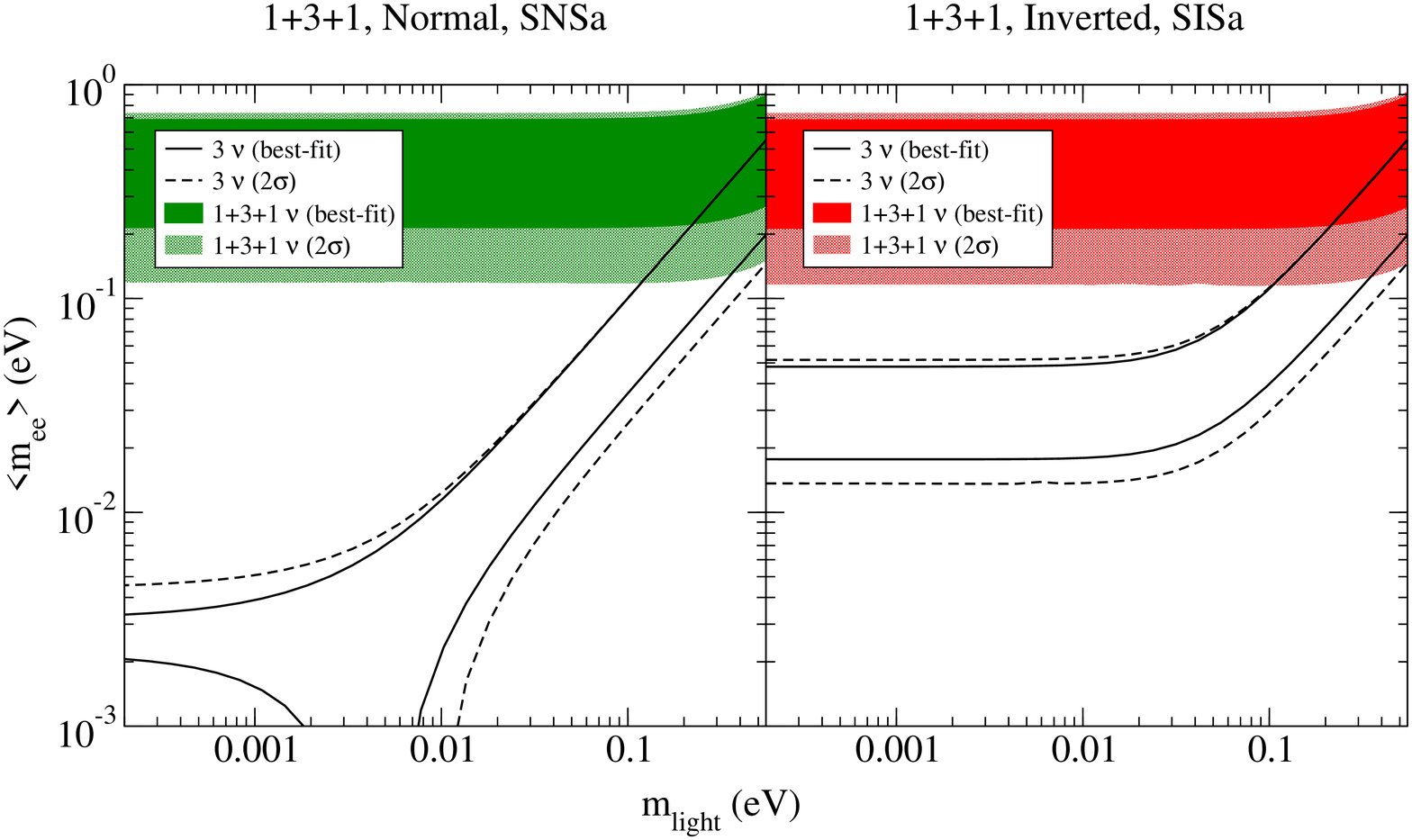}}
\subfigure{\label{fig:mee131b}
\includegraphics[width=0.8\textwidth]{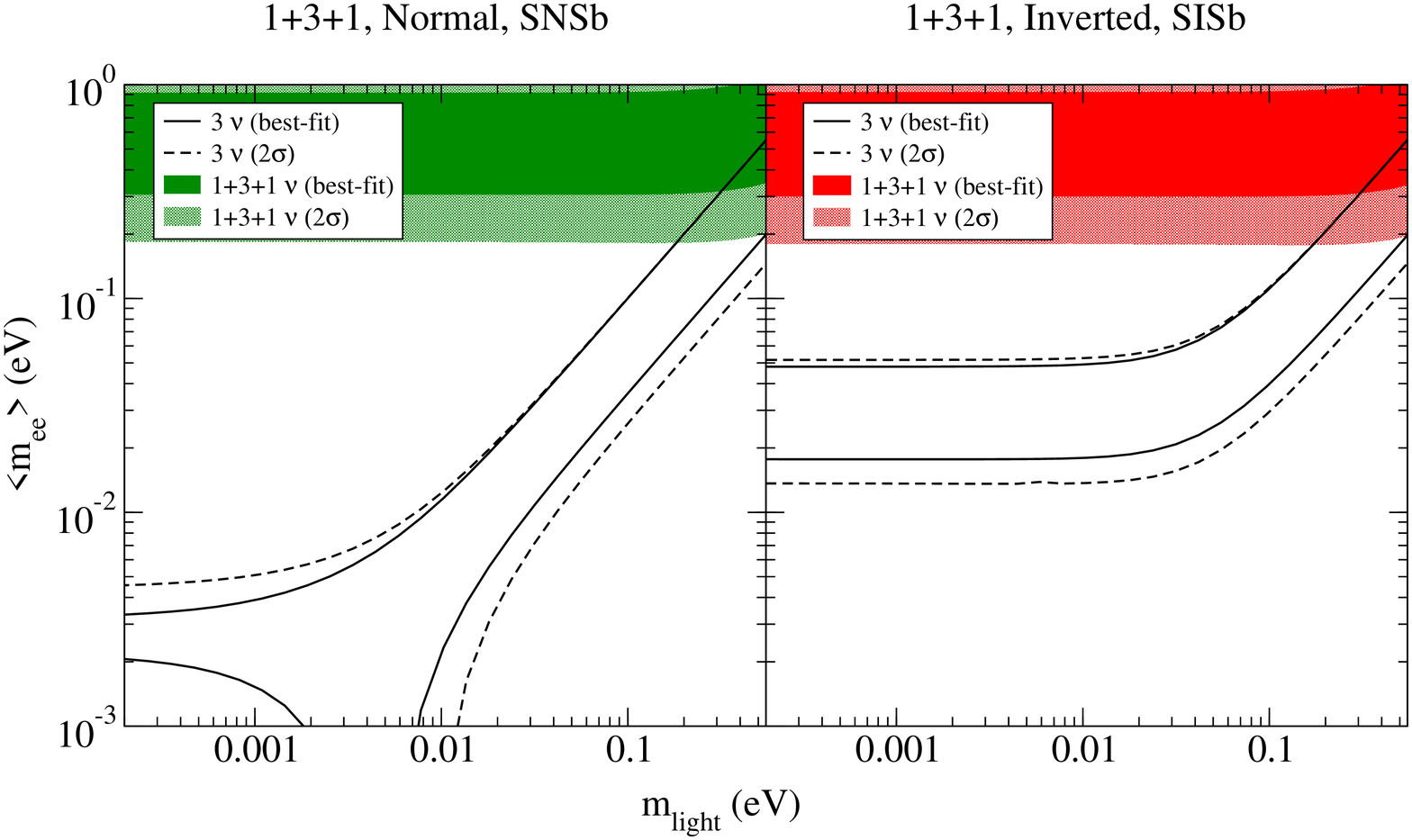}}
\caption{\label{fig:mee_5nu_b}Same as Fig.~\ref{fig:mee_4nu}, for the 
1+3+1a (top) and 1+3+1b (bottom) cases.}
\end{figure}

If there are two sterile neutrinos, the effective mass reads
\begin{equation}\label{eq:meff5}
\mee_{5\nu} = \left|c_{12}^2c_{13}^2c_{14}^2c_{15}^2m_1+
s_{12}^2c_{13}^2c_{14}^2c_{15}^2m_2e^{i\alpha}+s_{13}^2c_{14}^2c_{15}^2m_3e^{i\beta}
+s_{14}^2c_{15}^2m_4e^{i\gamma}+s_{15}^2m_5e^{i\phi}\right|  ,
\end{equation}
with $\phi$ the additional Majorana phase. In the 2+3 cases where
both of the sterile neutrinos are at the eV scale, $\mee$ can be approximated by
\begin{equation}
 \mee_{(2+3)\nu} \simeq  \left|c_{14}^2c_{15}^2\mee_{3\nu} + s_{14}^2\sqrt{\dmnewa}e^{i\gamma}+s_{15}^2\sqrt{\Delta
 m_{51}^2}e^{i\phi}\right|  ,
\label{eq:mee_3p2}
\end{equation}
in analogy to the 1+3 case. The upper panel of
Fig.~\ref{fig:mee_5nu_a} shows the allowed regions in this case: the
phenomenology is similar to that discussed for 1+3 above, except
that the presence of two sterile terms in Eq.~\eqref{eq:mee_3p2}
allows $\mee$ to take smaller values in the hierarchical region for
the normal ordering. The inverted ordering is essentially the same as in the 1+3 case. The two eV-scale sterile neutrinos dominate the KATRIN observable, with $\mbeta \gs \sqrt{|U_{e4}|^2 \, \dmnewa + |U_{e5}|^2 \, \dmnewb} \simeq 0.16$ eV, and the sum of masses is approximately $\sumnu \gs \sqrt{\dmnewa} + \sqrt{\dmnewb} \simeq 1.6$ eV.

The lower panel of Fig.~\ref{fig:mee_5nu_a} displays the 3+2 cases, where the sterile neutrinos are lighter than the active ones. Since the three active neutrinos are quasi-degenerate at the eV scale, with their mass given by the largest sterile mass-squared difference $\sqrt{\dmnewa} \simeq 0.69$ eV, Eq.~(\ref{eq:meffQD}) applies for $\mee_{(3+2)\nu}$ and the value of $\mbeta$ is also set by this scale. The mass ordering of the active states plays no role. Once again these scenarios are disfavored by cosmology, since $\sumnu \gs 3\sqrt{\dmnewb} + \sqrt{\dmnewb-\dmnewa} \simeq 3.4$ eV.

For completeness we include plots of the effective mass in the 1+3+1
schemes, where the active neutrinos are sandwiched between the sterile
ones. The plots in Fig.~\ref{fig:mee_5nu_b} show that there is little
difference between these scenarios and the 3+2 cases. The main
difference arises in SNSb and SISb scenarios, in which the effective
mass is governed by three quasi-degenerate neutrinos with mass
$\sqrt{\dmnewb} \simeq 0.93$ eV, plus a (coherent) contribution of 
$|U_{e4}|^2 \, \sqrt{\dmnewa + \dmnewb} \simeq 0.03$ eV. The KATRIN observable is also dominated by this mass scale, and the sum of masses is $\sumnu \gs 3\sqrt{\dmnewb} + \sqrt{\dmnewa+\dmnewb} \simeq 4.0$~eV.

The results of this section will not be significantly modified in the presence of new data~\cite{Abe:2011sj} on $\ssre$. In the hierarchical region with normal ordering, the lower limit on the effective mass decreases with $\ssre$, so that the larger best-fit value from Ref.~\cite{Fogli:2011qn} would have a small effect on $\mee$ in the standard three-neutrino scenario. However, in the sterile case the dominant terms in Eqs.~\eqref{eq:mee_3p1} and \eqref{eq:mee_3p2} are independent of $\ssre$.
\\


As discussed in the different cases above, if the predictions of the effective mass are above 0.2 eV, which is true for most of the possible scenarios, then KATRIN will see a signal as well. Indeed, Refs.~\cite{Riis:2010zm,Barrett:2011jg} show that sterile neutrinos with masses and mixings in the ranges considered above will be observable in KATRIN, and will be distinguishable from active neutrinos. This effect should be seen in almost all cases; the smallest contribution to the observable $\mbeta$ arises in the SSN and SSI cases.

In what regards the sum of masses, the smallest result holds again for SSN and SSI ($\sumnu \gs 1.6$ eV); the value of $\sumnu$ exceeds this in all other cases. The current upper bound from cosmology is approximately $\sumnu \ls 0.5$ eV \cite{Hannestad:2010kz}, which introduces tension with the sterile neutrino scenarios presented above. However, this measurement is highly model dependent and depends on the cosmological data set used. The Planck satellite should in the near future be able to reach a sensitivity of $\sim 0.1$ eV, thus providing more precise limits on the absolute neutrino mass. In addition, the present hints for extra radiation from analysis of CMB anisotropy measurements will be further constrained by Planck data, with a precision of $\Delta N_{\rm eff} \sim \pm 0.3$ \cite{Hamann:2007sb,Hamann:2010bk}, where $N_{\rm eff}$ is the effective number of thermally excited neutrino degrees of freedom. This means that Planck could in fact rule out a sterile neutrino species before beta decay or $\obb$ experiments see a signal.

\section{Sterile neutrinos in flavor symmetry models: an $A_4$ example}
\label{sec:A4}

In this section we modify a popular flavor symmetry model, which predicts tri-bimaximal mixing and is based on the group $A_4$, in order to accommodate one or two light sterile neutrinos. 

\subsection{One sterile neutrino}

In the Altarelli-Feruglio $A_4$ model from Ref.~\cite{Altarelli:2005yp}, neutrinos get mass from effective
operators, and the judicious choice of particle and flavon content along with the correct vacuum expectation value (VEV) alignment leads to TBM. The relevant particle assignments are shown in Table~\ref{table:afmodel}. Note the presence of an additional $Z_3$ symmetry to separate the neutrino and charged lepton sectors, and the Froggatt-Nielsen $U(1)_{\rm FN}$ to generate a hierarchy for the charged lepton masses. We have also included a
sterile neutrino $\nu_s$ with appropriate quantum numbers under the symmetries. 
\begin{table}[t]
\centering \caption{Particle assignments of the $A_4$ model, modified from Ref.~\cite{Altarelli:2005yp} to include a sterile
neutrino $\nu_s$. The additional $Z_3$ symmetry decouples the charged lepton and neutrino sectors; the $U(1)_{\rm FN}$ charge generates the hierarchy of charged lepton masses and regulates the scale of the sterile state.} 
\label{table:afmodel} \vspace{8pt}
\begin{tabular}{c|cccccccc|c}
  \hline \hline \T \B Field & $L$ & $e^c$ & $\mu^c$ & $\tau^c$ & $h_{u,d}$ & $\varphi$ & $\varphi'$ & $\xi$ & $\nu_s$ \\
\hline \T $SU(2)_L$ & $2$ & $1$ & $1$ & $1$ & $2$ & $1$ & $1$ & $1$ & $1$ \\
 $A_4$ & $\ul{3}$ & $\ul{1}$ & $\ul{1}''$ & $\ul{1}'$ & $\ul{1}$ & $\ul{3}$ & $\ul{3}$ & $\ul{1}$ & $\ul{1}$ \\
 $Z_3$ & $\omega$ & $\omega^2$ & $\omega^2$ & $\omega^2$ & 1 & 1 & $\omega$ & $\omega$ & 1 \\
 $U(1)_{FN}$ & - & 3 & 1 & 0 & - & - & - & - & 6 \\[1mm] \hline \hline
\end{tabular}
\end{table}

Leaving $\nu_s$ aside for the moment, these particle assignments, along with the $A_4$ multiplication
rules (see e.g. Refs.~\cite{Altarelli:2010gt,Ishimori:2010au}), lead to the Lagrangian
\begin{align}
  {\cal L}_{\rm Y} &= \ \frac{y_e}{\Lambda}e^c(\varphi L)h_d + \frac{y_{\mu}}{\Lambda}\mu^c(\varphi L)'h_d + \frac{y_{\tau}}{\Lambda}\tau^c(\varphi L)''h_d \notag \\[2mm]
    &+ \frac{x_a}{\Lambda^2}\xi(Lh_uLh_u) + \frac{x_d}{\Lambda^2}(\varphi'Lh_uLh_u) + {\rm h.c.} + \dots,
\label{eq:lag_AFmodel}
\end{align}
where $\Lambda$ is the cut-off scale 
and the dots stand for higher dimensional operators. The notation is
such that two fields $a$ and $b$ written as $(a b)$ transform as
$\ul{1}$, etc. If one chooses the real basis for $A_4$, along with the
flavon VEV alignments\footnote{Several methods have been proposed to explain
the different alignments and solve the so-called ''vacuum alignment
problem''; these will not be discussed here. Note that we do not
introduce more flavons to the original model, so that we can assume that
the solution of the alignment problem will not be modified.} $\langle \xi \rangle =
u$, $\langle \varphi \rangle = (v,0,0)$ and $\langle \varphi'
\rangle = (v',v',v')$, then the charged lepton mass matrix is
diagonal, and the neutrino mass matrix
\begin{equation}
M_\nu = \frac{v_u^2}{\Lambda}  
\begin{pmatrix} a'+\frac{2d'}{3} & -\frac{d'}{3} & -\frac{d'}{3} 
\\ \cdot & \frac{2d'}{3} & a'-\frac{d'}{3} 
\\ \cdot & \cdot & \frac{2d'}{3} \end{pmatrix} ,
 \label{eq:massm_af}
 \end{equation}
is diagonalized by the TBM matrix,
\begin{equation}
  U_{\rm TBM} = \begin{pmatrix} \frac{2}{\sqrt{6}} &
\frac{1}{\sqrt{3}} & 0 \\ -\frac{1}{\sqrt{6}} & \frac{1}{\sqrt{3}} &
-\frac{1}{\sqrt{2}} \\ -\frac{1}{\sqrt{6}} & \frac{1}{\sqrt{3}} &
\frac{1}{\sqrt{2}} \end{pmatrix} .
\label{eq:tbm}
\end{equation}
In this case $M_\nu$ is form-diagonalizable: the
eigenvectors are independent of the parameters in the neutrino mass
matrix. The charged lepton mass hierarchy is generated by the
Froggatt-Nielsen (FN) mechanism: the $U(1)_{\rm FN}$ charges 0, 1 and 
3 are assigned to the right-handed charged singlets $\tau^c$,
$\mu^c$ and $e^c$, respectively, and a flavon $\theta$ that carries a negative unit of this charge is introduced, suppressing each mass term by powers of the small parameter $\langle \theta \rangle/\Lambda \equiv \lambda < 1$. Explicitly,
\begin{equation}
 m_\alpha = y_\alpha v_d\frac{v}{\Lambda} \lambda^{F_\alpha}\, ,
\end{equation}
where $F_\alpha$ is the relevant $U(1)_{\rm FN}$ charge.

By assuming that (i) the Yukawa couplings $y_\alpha$, $x_a$ and
$x_d$ remain in a perturbative regime; (ii) the flavon VEVs are
smaller than the cut-off scale and (iii) all flavon VEVs fall in
approximately the same range, the authors of
Ref.~\cite{Altarelli:2005yp} obtain the relation
\begin{equation}
 0.004 < \frac{v'}{\Lambda} \approx \frac{v}{\Lambda} \approx \frac{u}{\Lambda} < 1\, ,
\label{eq:flavonscales}
\end{equation}
with the cut-off scale $\Lambda$ ranging between $10^{12}$ and
$10^{15}$ GeV, and $v_u \approx 174$ GeV.\\

In order to accommodate sterile neutrinos in this framework, we have 
added an additional sterile singlet, $\nu_s$, transforming as
$\ul{1}$ under $A_4$ and $1$ under $Z_3$. The $A_4$ invariant dimension-5 operator $\frac{1}{\Lambda}(\varphi' L h_u)\nu_s$ is not allowed by the $Z_3$ symmetry, and we
are left with the terms
\begin{equation}
 {\cal L}_{{\rm Y}_s} = \frac{x_e}{\Lambda^2}\xi(\varphi' Lh_u)\nu_s + \frac{x_f}{\Lambda^2}(\varphi'\varphi' Lh_u)\nu_s + m_s\nu^c_s\nu^c_s + {\rm h.c.},
\label{eq:lagsterile}
\end{equation}
where $m_s$ is a bare Majorana mass. As we will see, the chosen FN
charge forces $m_s$ to be at the desired eV scale. 
The modified $4\times 4$ mass matrix is (cf. Eq.~\eqref{eq:massm_af}) 
\begin{equation}
 M^{4\times4}_\nu = \begin{pmatrix} a+\frac{2d}{3} & -\frac{d}{3} & -\frac{d}{3} & e \\ \cdot & \frac{2d}{3} & a-\frac{d}{3} & e \\ \cdot & \cdot & \frac{2d}{3} & e \\ \cdot & \cdot & \cdot & m_s \end{pmatrix},
\label{eq:m4by4}
\end{equation}
where $a = 2x_a\frac{u v_u^2}{\Lambda^2}$, $d = 2x_d\frac{v'
v_u^2}{\Lambda^2}$ and $e = \sqrt{2}x_e \frac{u v'v_u}{\Lambda^2}$
have dimensions of mass. Note that the first three elements of the
fourth row of $M^{4\times4}_\nu$ are identical because of the VEV
alignment $\langle \varphi' \rangle = (v',v',v')$, which was necessary
to generate TBM in the 3 neutrino case; this alignment combined with the $A_4$ multiplication rules also causes the second term in Eq.~\eqref{eq:lagsterile} (proportional to $x_f$) to vanish.

The correct mass scales can be obtained by assigning a $U(1)_{\rm FN}$
charge of $F_{\nu_s} = 6$ to $\nu_s$, in analogy to the mechanism
used in the charged lepton sector. Indeed, in order to fit the data,
the parameters $a$ and $d$ should be between $10^{-3}$ and $10^{-1}$
eV, the ratio $e/m_s \sim {\cal O}(10^{-1})$ to generate sufficient
mixing, and $m_s \sim 1$ eV for the sterile neutrino mass.

As an explicit example, assume that $v'/\Lambda \approx v/\Lambda
\approx u/\Lambda \simeq 10^{-1.5}$, in keeping with the constraints
from Eq.~\eqref{eq:flavonscales}, and that the cut-off scale is
$\Lambda \simeq 10^{12.5}$ GeV. In this case one obtains
\bea \D 
 a \sim d \simeq 0.1 \left(\frac{u}{10^{11}\ {\rm
GeV}}\right)\left(\frac{v_u}{10^{2} \ {\rm GeV}}\right)^2
\left(\frac{10^{12.5}\ {\rm GeV}}{\Lambda}\right)^2 \ {\rm eV}\, ,
\\[2mm] \D 
 e \simeq 0.1 \left(\frac{\lambda}{10^{-1.5}}\right)^6\left(\frac{u}{10^{11}\ {\rm GeV}}\right)\left(\frac{v'}{10^{11}\ {\rm GeV}}\right)\left(\frac{v_u}{10^{2} \ {\rm GeV}}\right) \left(\frac{10^{12.5}\ {\rm GeV}}{\Lambda}\right)^2 \ {\rm eV}\, ,
\eea
with the assumption that the Yukawa couplings $x_{a,d,e}$ are of
order 1, and that $\lambda \approx 10^{-1.5}$, so that $\langle
\theta \rangle$ is in the same range as the other flavon VEVs.

The Majorana mass term $m_s\nu^c_s\nu^c_s$ is doubly suppressed by the $U(1)_{\rm FN}$ charge, and there are additional terms which can give a contribution to this mass in addition to the bare term. From the particle assignments in
Table~\ref{table:afmodel}, the leading and next-to-leading order non-vanishing contributions to $m_s$ are\footnote{The term proportional to $\frac{x_{s'''}}{\Lambda^2}(\varphi'\varphi'\varphi')$ vanishes after $A_4$ symmetry breaking.}
\begin{equation}
 \left(\frac{x_s}{\Lambda}(\varphi\varphi)+\frac{x_{s'}}{\Lambda^2}\xi\xi\xi + \frac{x_{s''}}{\Lambda^2}(\varphi'\varphi')\xi \right)\nu^c_s\nu^c_s \Longrightarrow \left({x_s}\frac{v^2}{\Lambda} + x_{s'}\frac{u^3}{\Lambda^2} + x_{s''}\frac{3v'^2 u}{\Lambda^2}\right)\lambda^{2F_\nu}\, ,
\label{eq:majcontrib}
\end{equation}
so that these terms are suppressed by $\lambda^{12}$, and the
resulting Majorana mass can be of order eV, i.e. 
\begin{equation}
 m_s \simeq 10^{0.5} \left(\frac{\lambda}{10^{-1.5}} \right)^{12} \left(\frac{v}{10^{11}\ {\rm GeV}}\right)^2 \left(\frac{10^{12.5}\ {\rm GeV}}{\Lambda}\right) \ {\rm eV}\, .
\end{equation}
In this case the contributions from the second and third terms in
Eq.~\eqref{eq:majcontrib} are of order 0.1~eV, and do not affect the
scale of $m_s$ significantly. We conclude that the usual choice of
scales and charges easily allows $a \sim d \sim e < m_s$ in
$M^{4\times4}_\nu$ from Eq.~(\ref{eq:m4by4}). 

\begin{figure}[t]
 \centering
 \includegraphics[width=0.6\textwidth]{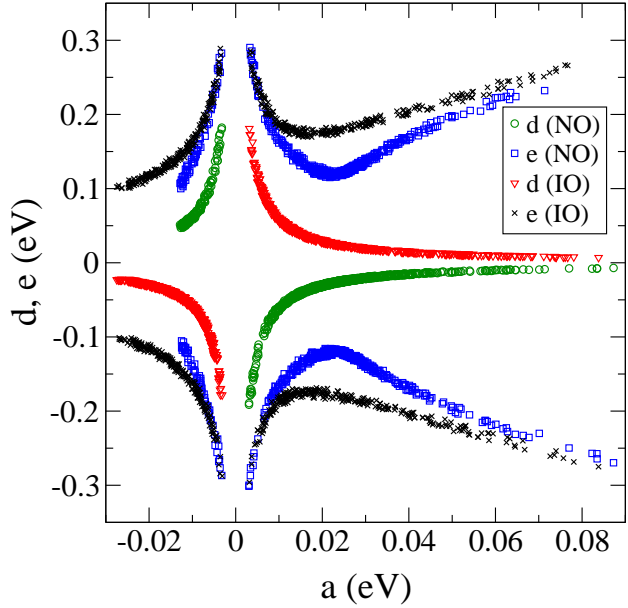}
 \caption{The allowed values in $a-d$ and $a-e$ parameter space for
normal (NO) and inverted (IO) ordering, obtained by varying each
parameter between $-0.5$ and $0.5$ eV, varying $m_s$ between $-1.5$
and $1.5$ eV, and requiring that the oscillation parameters lie in the correct range \cite{Kopp:2011qd,Schwetz:2011qt}.
\label{fig:ade}}
\end{figure}

Assuming that the parameters are all real, Eq.~(\ref{eq:m4by4}) is exactly diagonalized by
 \begin{equation}
  U = \begin{pmatrix} \frac{2}{\sqrt{6}} & \frac{1}{6e}\frac{K_-}{N_-} & 0 & \frac{1}{6e}\frac{K_+}{N_+} \\[2mm] -\frac{1}{\sqrt{6}} &  \frac{1}{6e}\frac{K_-}{N_-} & -\frac{1}{\sqrt{2}} & \frac{1}{6e}\frac{K_+}{N_+} \\[2mm] -\frac{1}{\sqrt{6}} & \frac{1}{6e}\frac{K_-}{N_-} & \frac{1}{\sqrt{2}} & \frac{1}{6e}\frac{K_+}{N_+} \\[2mm] 0 & \frac{1}{N_-} & 0 & \frac{1}{N_+} \end{pmatrix}\, ,
 \label{eq:v4_ex}
\end{equation}
where $K_{\pm}=a-m_s\pm\sqrt{12e^2+(a-m_s)^2}$ and $N_{\pm}^2=1+\frac{\left(a-m_s \pm \sqrt{12 e^2+(a-m_s)^2}\right)^2}{12 e^2}$. 
If one assumes that $a < m_s$ and expands to second order in the small ratio $e/m_s$, the resulting mixing matrix is
\begin{equation}
   U \simeq \begin{pmatrix} \frac{2}{\sqrt{6}} & \frac{1}{\sqrt{3}} & 0 & 0 \\ -\frac{1}{\sqrt{6}} & \frac{1}{\sqrt{3}} & -\frac{1}{\sqrt{2}} & 0 \\ -\frac{1}{\sqrt{6}} & \frac{1}{\sqrt{3}} & \frac{1}{\sqrt{2}} & 0 \\ 0 & 0 & 0 & 1 \end{pmatrix} + \begin{pmatrix} 0 & 0 & 0 & \frac{e}{m_s} \\ 0 & 0 & 0 & \frac{e}{m_s} \\ 0 & 0 & 0 & \frac{e}{m_s} \\ 0 & -\frac{\sqrt{3}e}{m_s} & 0 & 0 \end{pmatrix} + \begin{pmatrix} 0 & -\frac{\sqrt{3} e^2}{2m_s^2} & 0 & 0 \\ 0 & -\frac{\sqrt{3} e^2}{2m_s^2} & 0 & 0 \\ 0 & -\frac{\sqrt{3} e^2}{2 m_s^2} & 0 & 0 \\ 0 & 0 & 0 & -\frac{3e^2}{2 m_s^2}\end{pmatrix}  ,
\label{eq:v4}
\end{equation}
giving the eigenvalues
\be
 m_1 = a+d\, ,~~  m_2 = a - \frac{3e^2}{m_s}\, , 
~~  m_3 = -a+d\, ,  ~~m_4 = m_s + \frac{3e^2}{m_s}\, . 
\ee 
In this case one can see that $M^{4\times4}_\nu$ is not form-diagonalizable anymore: the second and fourth column of $U$ are sensitive to the entries of $M^{4\times4}_\nu$. 
The mass-squared differences as well
as the active-sterile mixing angles $\sin^2 \theta_{i4}$
($i=1,2,3$) are controlled by the four parameters in 
Eq.~(\ref{eq:m4by4}), with the mixing most sensitive to the ratio
$e/m_s$. Note that both the normal and inverted orderings are
allowed, in contrast to the standard three neutrino version of the
model, which only allowed the normal ordering. By fitting the parameters
$a$, $d$, $e$ and $m_s$ to the allowed range of the four parameters $\dms$, $\dma$, $\dmnewa$ and
$\sssta$, we find that the masses are arbitrary, are not constrained to any particular region, and are in general uncorrelated with the mixing parameters. However, the lightest mass increases with $\sssta$, which means that the effective mass in $\obb$, $\mee \equiv \left|a+\frac{2d}{3}\right|$, also increases with $\sssta$, as expected from Eq.~\eqref{eq:mee_3p1}. Figure~\ref{fig:ade} shows the allowed ranges of $a,d,e$. In general, $a$ and $d$ are
approximately inversely proportional to each other. 

\begin{figure}[t]
 \centering
 \includegraphics[angle=270,width=0.8\textwidth]{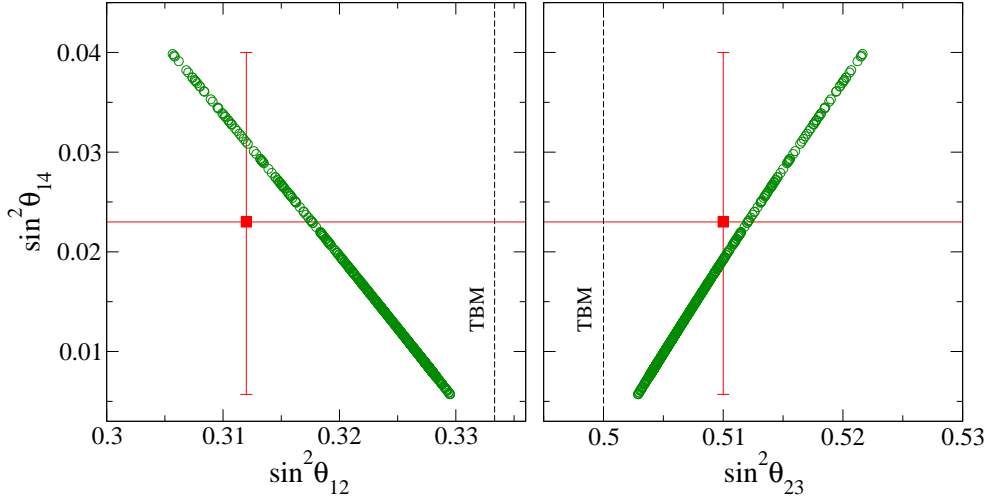}
 \caption{$\sssta$ against $\sssol$ and $\ssatm$, for both the normal
and inverted ordering. The dashed (black) lines corresponds to the TBM
values of $\sssol$ and $\ssatm$, the solid (red) lines indicate the
$2\sigma$ ranges of the parameters and the (red) square is the
best-fit point \cite{Kopp:2011qd,Schwetz:2011qt}.}
\label{fig:sinsq1214}
\end{figure}

Comparison of Eqs.~\eqref{eq:UPMNS4x4} and \eqref{eq:v4}, shows that
$\sin \theta_{13}=0$, i.e.~this parameter retains its TBM value, whereas
$\sssol$ and $\ssatm$ receive small corrections:
\bea \D
 \sssol = \frac{|U_{e2}|^2}{1-|U_{e4}|^2} \simeq \frac{1}{3}\left[1 - 2\left(\frac{e}{m_s}\right)^2\right]\, , \\[2mm]\D
 \ssatm = \frac{|U_{\mu3}|^2(1-|U_{e4}|^2)}{1-|U_{e4}|^2-|U_{\mu4}|^2} \simeq \frac{1}{2}\left[1 + \left(\frac{e}{m_s}\right)^2\right]\, .\D
\label{eq:sinsq12_23}
\eea
Note the correlation $\ssatm \simeq \frac 34 (1 - \sssol)$ following from
the above expressions. Other results of the model are $U_{s1} = U_{s
3} = 0$ and $U_{e4} =  U_{\mu4} = U_{\tau 4}$. The three active-sterile mixing angles 
can be expressed in terms of the matrix elements $U_{f4}$ as
\begin{equation}
 \sssta = |U_{e4}|^2\, , \quad \ssstb = \frac{|U_{\mu4}|^2}{1-|U_{e4}|^2}\, , \quad
 \ssstc = \frac{|U_{\tau4}|^2}{1-|U_{e4}|^2-|U_{\mu4}|^2}\, ,
\end{equation}
and the model predicts them all to be of similar magnitude:
\begin{equation}
 \sssta \simeq \ssstb \simeq \ssstc \simeq
\left(\frac{e}{m_s}\right)^2 \simeq 
\frac 12 (1 - 3\sssol ) \simeq 2 \ssatm - 1 \, ,
\label{eq:sinsq_i4}
\end{equation}
to second order in the ratio $e/m_s$. The correlation (see
Eqs.~\eqref{eq:sinsq12_23} and \eqref{eq:sinsq_i4}) between the
solar and atmospheric mixing parameters ($\sssol$ and $\ssatm$) and
the active-sterile mixing parameter ($\sssta$) is shown in
Fig.~\ref{fig:sinsq1214}, and is the same for both mass orderings. 

Although this model appears to predict $\theta_{13}=0$, this result depends on the triplet VEV alignments in the scalar sector. In the general case \cite{Honda:2008rs,Barry:2010zk}, these alignments could be modified to $\langle \varphi \rangle = (v,v\,\epsilon_1^{\rm ch},v\,\epsilon_2^{\rm ch})$ and $\langle \varphi' \rangle = (v',v'(1+\epsilon_1),v'(1+\epsilon_2))$, where the deviation parameters come from higher dimensional operators or renormalization group effects. Mixing angles will then receive corrections of the same order, proportional to $\epsilon_{1,2}^{\rm ch}$ and $\epsilon_{1,2}$, so that one can accommodate the latest T2K result as well as the allowed range from Ref.~\cite{Fogli:2011qn}. These perturbations would also affect the fourth column of Eq.~\eqref{eq:m4by4}, thus altering the results in Eqs.~\eqref{eq:sinsq12_23} and \eqref{eq:sinsq_i4}.

We note here that the small mixing with sterile neutrinos will in general modify mixing scenarios. The mixing angles of active and sterile neutrinos are of order $e/m_s$, where $e$ is any of the entries $(M_\nu^{4\times4})_{fs}$ with $f = e,\mu,\tau$. Deviations from initial mixing angles $\theta_{12,13,23}$ are then of the same order. The example we have discussed here has the particular feature that $U_{e4} =  U_{\mu4} = U_{\tau 4}$ and that the first and third row of $U$ are identical to TBM. This will be different in other cases. 


In a more general flavor symmetry context, it is instuctive to note the origin of the eigenvector proportional to $(0,-1,1,0)^T$ in Eq.~(\ref{eq:v4}). This arises because of a generalized $\mu$-$\tau$ symmetry. Consider an
arbitrary $4\times4$ Majorana mass matrix $M_\nu^{4 \times 4}$. The defining matrix $P_{\mu \tau}$ for the $Z_2$ corresponding to $\mu$-$\tau$ symmetry fulfills $P_{\mu \tau}^2 = \mathbbm{1}$ and the invariance condition $P_{\mu \tau} M_\nu^{4 \times 4} P_{\mu \tau} = M_\nu^{4 \times 4}$, so that $P_{\mu \tau}$ and the resulting mass matrix are given by
\be
P_{\mu \tau} = 
\left( \baf 
1 & 0 & 0 & 0 \\
0 & 0 & 1 & 0 \\
0 & 1 & 0 & 0 \\
0 & 0 & 0 & 1 
\ea \right) \quad \mbox{ and } \quad
M_\nu^{4 \times 4} = \left( \baf 
\tilde a & \tilde b & \tilde b & \tilde d \\
\cdot & \tilde e & \tilde f & \tilde g \\
\cdot & \cdot & \tilde e & \tilde g \\
\cdot & \cdot & \cdot & \tilde m
\ea
\right) .
\ee
The eigenvalue $\tilde e-\tilde f$ of this mass matrix corresponds to an eigenvector proportional to $(0,-1,1,0)^T$. This $Z_2$ invariance, usually present in $A_4$ models \cite{Altarelli:2010gt}, arises by spontaneous $A_4$ breaking. 
Regarding the nearly TBM mixing, there is a second $Z_2$ under which $M^{4\times4}_\nu$ is invariant. It is defined by the generator 
\be
P_{\rm sol} = \frac 13 \left( \baf 
-1 & 2 & 2 & 0 \\
2 & -1 & 2 & 0 \\
2 & 2 & -1 &  0 \\
0 & 0 & 0 & 3 
\ea \right) . 
\ee
Additional invariance under this $Z_2$ (note that $P_{\mu \tau}$ and $P_{\rm sol}$ commute) requires $\tilde f = \tilde a - \tilde e + \tilde b$ and $\tilde d = \tilde g$. As a result, the eigenvalue $2\tilde e - \tilde b - \tilde a$ has an eigenvector proportional to $(-2,1,1,0)^T$. Furthermore, it holds that $U_{e s} = U_{\mu s} = U_{\tau s}$. The main features of Eq.~(\ref{eq:m4by4}) are therefore explained by the two $Z_2$ symmetries. In addition, the simplicity of the $A_4$ model leads to $\tilde e = -2\tilde b$, but does not modify the mixing properties that arise from the  two $Z_2$ symmetries.

Let us denote with $S$ the upper left $3\times3$ part of $P_{\rm
sol}$. With a suitably chosen diagonal phase matrix $T = {\rm diag}(1,e^{-i
2\pi/3},e^{i 2\pi/3})$, under which the charged lepton mass matrix 
$M_\ell M_\ell^\dagger$ is invariant, it holds that $T^3 = (S T)^3 =
\mathbbm{1}$. Hence, $S$ and $T$ generate $A_4$, and the charged
lepton $Z_3$ defined by $T$ arises again by spontaneous $A_4$
breaking. These well known features \cite{Altarelli:2010gt} of $A_4$
models are not altered by our modification. 

\subsection{Two sterile neutrinos}

In order to have two sterile neutrinos in the $A_4$ model, one can
simply add a second sterile singlet $\nu_{s_2}$. As before, this
sterile neutrino is a singlet $\ul{1}$ under $A_4$ and 1 under $Z_3$, and carries the
$U(1)_{\rm FN}$ charge $F_{\nu_s} = 6$. With the additional assumption
that the sterile sector of the mass matrix is
diagonal\footnote{This can be achieved, for example, with an
additional discrete symmetry such as $Z_2$ operating only in the
sterile sector.}, the symmetric $5\times5$ mass matrix is
\begin{equation}
 M^{5\times5}_\nu = \begin{pmatrix} a+\frac{2d}{3} & -\frac{d}{3} & -\frac{d}{3} & e & f \\ \cdot & \frac{2d}{3} & a-\frac{d}{3} & e & f \\ \cdot & \cdot & \frac{2d}{3} & e & f \\ \cdot & \cdot & \cdot & m_{s_1} & 0 \\ \cdot & \cdot & \cdot & \cdot & m_{s_2} \end{pmatrix}.
\label{eq:m5by5}
\end{equation}
Similar statements about the $Z_2$ invariance of properly extended $P_{\mu \tau}$ and
$P_{\rm sol}$ symmetries can be made here (see the discussion at the
end of the last subsection). Since $f$ and $m_{s_2}$ arise in analogy to $e$ and  $m_{s}$ in the
one sterile neutrino case, we expect that $e$ and $f$, as well as
$m_{s_2}$ and $m_{s_2}$, are each of similar magnitude, respectively. In analogy to the case discussed in the previous subsection, the mass matrix is approximately diagonalized by
\bea \D
   U = \begin{pmatrix} \frac{2}{\sqrt{6}} & \frac{1}{\sqrt{3}} & 0 & 0 & 0 \\ -\frac{1}{\sqrt{6}} & \frac{1}{\sqrt{3}} & -\frac{1}{\sqrt{2}} & 0 & 0 \\ -\frac{1}{\sqrt{6}} & \frac{1}{\sqrt{3}} & \frac{1}{\sqrt{2}} & 0 & 0 \\ 0 & 0 & 0 & 1 & 0 \\ 0 & 0 & 0 & 0 & 1 \end{pmatrix} + \begin{pmatrix}0 & 0 & 0 & \frac{e}{m_{s_1}} & \frac{f}{m_{s_2}} \\ 0 & 0 & 0 & \frac{e}{m_{s_1}} & \frac{f}{m_{s_2}} \\ 0 & 0 & 0 & \frac{e}{m_{s_1}} & \frac{f}{m_{s_2}} \\ 0 & -\frac{\sqrt{3}e}{m_{s_1}} & 0 & 0 & 0 \\  0 & -\frac{\sqrt{3}f}{m_{s_2}} & 0 & 0 & 0 \end{pmatrix} \ \\[3mm] \D + \begin{pmatrix} 0 & -\frac{\sqrt{3}}{2}\left(\frac{e^2}{m_{s_1}^2}+\frac{f^2}{m_{s_2}^2}\right) & 0 & 0 & 0 \\  0 & -\frac{\sqrt{3}}{2}\left(\frac{e^2}{m_{s_1}^2}+\frac{f^2}{m_{s_2}^2}\right) & 0 & 0 & 0 \\ 0 & -\frac{\sqrt{3}}{2}\left(\frac{e^2}{m_{s_1}^2}+\frac{f^2}{m_{s_2}^2}\right) & 0 & 0 & 0 \\ 0 & 0 & 0 & -\frac{3e^2}{2m_{s_1}^2} & -\frac{3ef}{2m_{s_1}m_{s_2}} \\ 0 & 0 & 0 & -\frac{3ef}{2m_{s_1}m_{s_2}} & -\frac{3f^2}{2m_{s_2}^2} \end{pmatrix} \, , \D
\label{eq:v5}
\eea
assuming that $a<m_{s_{1,2}}$ and that the ratios $e/m_{s_1}$ and
$f/m_{s_2}$ are small. The mass eigenvalues are
\bea 
 m_1 = a+d\, ,~~  m_2 = a - \dfrac{3e^2}{m_{s_1}} -
\dfrac{3f^2}{m_{s_2}}\, , ~~ m_3 = -a+d\, , ~~ m_4 = m_{s_1} + \dfrac{3e^2}{m_{s_1}}\, , ~~
 m_5 = m_{s_2} + \dfrac{3f^2}{m_{s_2}}\, , 
\eea
to second order in the ratios $e/m_{s_1}$ and $f/m_{s_2}$. 

Once again, the reactor mixing angle retains its TBM value, and the
predictions for $\sssol$ and $\ssatm$ are
\bea \D 
  \sssol \simeq \frac{1}{3}\left[1 - 2\left(
\left(\frac{e}{m_{s_1}}\right)^2 +
\left(\frac{f}{m_{s_2}}\right)^2\right)\right]\, , \\[2mm] \D 
  \ssatm  \simeq \frac{1}{2}\left[1 + \left(\frac{e}{m_{s_1}}\right)^2+\left(\frac{f}{m_{s_2}}\right)^2\right]\, ,
 \label{eq:sinsq12_23_2}
\eea 
in analogy to the $4\times4$ case. Using the explicit
parameterization (\ref{eq:U2st}) of the $5\times5$ mixing matrix, the six
active-sterile mixing angles can be approximated by
\begin{equation}
 \sin^2\!\theta_{i4} \simeq \left(\frac{e}{m_{s_1}}\right)^2\, , \quad \sin^2\!\theta_{i5} \simeq \left(\frac{f}{m_{s_2}}\right)^2 \quad (i=1,2,3)\, ,
\label{eq:sinsq_i4_i5}
\end{equation}
with the additional assumption that $U$ is real.

Note that without an additional discrete symmetry the sterile sector
of Eq.~\eqref{eq:m5by5} would be a $2\times2$ democratic matrix with entries of
order $m_{s_1} \simeq m_{s_2} \simeq 1$ eV. In this case the
active-sterile mixing would be modified by the presence of an 
additional large 4-5 rotation in the overall $5\times5$ mixing matrix.

\section{Realization of light sterile neutrinos in seesaw models}
\label{sec:model}

We now discuss how one can accommodate eV-scale sterile neutrinos in seesaw frameworks. We start from the simplest type I seesaw scenario and then present an interesting extension to this.

\subsection{Sterile neutrinos from type I seesaw}

In the canonical type I seesaw model, three heavy right-handed neutrinos \mbox{$\nu_R=(\nu_{R1},\nu_{R2},\nu_{R3})$} are introduced and the neutrino mass Lagrangian reads
\begin{eqnarray}\label{eq:L1}
-{\cal L}_m = \overline{\nu_L} M_D \nu_R  +
\frac{1}{2}\overline{\nu^c_R} M_R \nu_R + {\rm h.c.},
\end{eqnarray}
where the Dirac mass matrix $M_D$ is an arbitrary matrix, while $M_R$ is symmetric according
to the Majorana nature of right-handed neutrinos. In the basis
$(\nu_L,\nu^c_R)$, the neutrino mass matrix is a $6\times 6$
matrix: 
\begin{eqnarray} \label{eq:Mnu}
M_\nu = \left(\begin{matrix} 0 & M_D \cr M^T_D & M_R \end{matrix}\, 
\right)\, ,
\end{eqnarray}
and if the entries of $M_D$ are all much smaller than the eigenvalues of
$M_R$, light neutrinos acquire masses after the heavy right-handed neutrinos
are integrated out, {\em viz.}
\begin{eqnarray}\label{eq:m-type-I}
m_\nu \simeq - M_D M^{-1}_R M^T_D   \, .
\end{eqnarray}
The mixing between heavy and light neutrinos comes from the full
diagonalization of $M_\nu$, and is approximately given by 
\begin{eqnarray}\label{eq:R}
R \simeq M_D M^{-1}_R  \, ,
\end{eqnarray}
in the basis where $M_R$ is diagonal.
The non-unitary mixing matrix relating the light neutrino 
flavor eigenstates and their mass eigenstates is modified to
\begin{eqnarray}
V \simeq (1-\frac{1}{2}RR^\dagger) U \, ,
\end{eqnarray}
where $U$ is a unitary matrix satisfying $U^\dagger m_\nu U^* ={\rm
diag}(m_1,m_2,m_3)$. Thus, in principle the type I seesaw mechanism
predicts sterile neutrinos mixed with active neutrinos. 
However, with $M_D$ naturally located at the electroweak
scale of $10^2~{\rm GeV}$ and from the requirement of sub-eV light neutrino
masses, one typically has that $M_R \sim 10^{14}~{\rm GeV}$. The
mass and mixing parameters are therefore not suited to explain eV-scale light neutrinos. Possible ways out of this dilemma are (i) to lower the overall  seesaw scale down to $M_R \sim $ eV
\cite{deGouvea:2006gz}; or 
(ii) bring one of the heavy neutrinos and the Dirac mass matrix
entries associated with it down to the eV scale. This is in the spirit of
the $\nu$MSM \cite{Asaka:2005an}, where however the adjective
``light'' denotes keV sterile neutrinos, which can act as warm dark
matter particles. 

In the first case, one may use for instance $M_R \simeq 1.5 ~{\rm
eV}$ and $M_D \simeq 0.3 ~{\rm eV}$, for which the light neutrino mass scale
lies around $0.06 ~{\rm eV}$ and $R \simeq 0.2$ according to
Eqs.~\eqref{eq:m-type-I} and \eqref{eq:R}. This is roughly in
agreement with the recently reported range of sterile neutrinos from
the reactor anti-neutrino anomaly. One would expect three sterile
neutrinos in this case. 
In this scenario, there is no
neutrino-less double beta decay, since both the
active and sterile neutrinos contribute to the effective mass
$\mee$, and their contributions are exactly cancelled
\cite{deGouvea:2006gz}. This is a consequence of the zero in the upper
left entry of Eq.~(\ref{eq:Mnu}) and the fact that all six neutrino
masses are below 100 MeV, which is the typical momentum exchange in
$\obb$. In addition, neither the baryon 
asymmetry problem nor the dark matter problem is solved in this 
framework, since the sterile neutrinos are
too light. It is also questionable why $M_D$ should be so tiny compared to
the weak scale, since both arise from Yukawa couplings to the
SM Higgs doublet.\\ 

Consider now the second possibility, namely bringing one right-handed neutrino down
to the eV scale. In order to simultaneously explain the baryon asymmetry of the
Universe and the reactor anomaly, a minimal extension of the SM
requires at least one light right-handed (sterile) neutrino $\nu_s$
together with two heavy right-handed neutrinos ($\nu_{R1}$ and
$\nu_{R2}$). In this case, the Lagrangian for neutrino masses is
\begin{eqnarray}\label{eq:L2}
-{\cal L}_m = \overline{\nu_L} M_D \nu_R  + \overline{\nu_L} M_S
\nu_s + \frac{1}{2}\overline{\nu^c_R} M_R \nu_R +
\frac{1}{2}\overline{\nu^c_S} \mu_S \nu_S  +  {\rm h.c.},
\end{eqnarray}
where $M_D$, $M_R$ and $M_S$ are $3\times 2$, $2\times 2$ and $3
\times 1$ mass matrices, respectively. 
The full neutrino mass matrix in the basis $(\nu_L, \nu^c_s, \nu^c_R)$ reads
\begin{eqnarray}
M_\nu = \left(\begin{matrix} 0 & 0 & 0 & (M_S)_{11} & (M_D)_{11} &
(M_D)_{12} \cr 0 & 0 & 0 & (M_S)_{21} & (M_D)_{21} & (M_D)_{22} \cr
0 & 0 & 0 & (M_S)_{31} & (M_D)_{31} & (M_D)_{32} \cr (M_S)_{11} &
(M_S)_{21} & (M_S)_{31} & \mu_S & 0 & 0 \cr (M_D)_{11} & (M_D)_{21}
& (M_D)_{31} & 0 & M_1 & 0 \cr (M_D)_{12} & (M_D)_{22} & (M_D)_{32}
& 0 & 0 & M_2
\end{matrix} \right) ,
\end{eqnarray}
where we work in the basis where $M_R$ is diagonal (this is
always possible by performing a proper basis transformation).

We assume that two of the right-handed neutrinos are heavy (e.g.~above
the TeV scale or higher) and $\mu_S$ is much smaller than $M_R$,
i.e. around the eV scale. Then, at energy scales much lower than $M_R$,
$\nu_{Ri}$ should be decoupled 
from the full theory, and one obtains a $4\times 4$ matrix for active and sterile neutrinos, {\em viz.}
\begin{eqnarray}\label{eq:M4}
M_\nu^{4\times 4}  = \left(\begin{matrix} -M_D M^{-1}_R M^T_D &
M_S \cr M^T_S & \mu_S
\end{matrix} \right) .
\end{eqnarray}
In the case where $\mu_S \gg M_S$, one may apply the seesaw formula
once again, and arrive at the light neutrino mass matrix
\begin{eqnarray}
m_\nu  \simeq -M_D M^{-1}_R M^T_D - M_S \mu^{-1}_S M^T_S \, .
\end{eqnarray}
The mixing between active and sterile neutrinos is characterized by $R=M_S \mu^{-1}_S$. Comparison with
Eq.~\eqref{eq:UPMNS4x4} gives
\begin{eqnarray}
\sin\theta_{14} & \simeq & (M_S)_{11}\mu^{-1}_S \, ,
\end{eqnarray}
together with the mass of the sterile neutrino
\begin{eqnarray}
m_s & \simeq &  \mu_S \, .
\end{eqnarray}
The scenario under discussion permits neutrino-less double beta decay, since the active and sterile neutrino contributions do not cancel with each other. Furthermore, in this scenario the baryon asymmetry of the Universe can be explained via the leptogenesis mechanism, for which at least two heavy right-handed neutrinos are required. However, it should also be noticed that none of the singlet neutrinos have masses in the allowed range of sterile neutrino warm dark matter (i.e.~around ${\rm keV}$), so that one cannot simultaneously solve the dark matter puzzle~\cite{Asaka:2005an}. 


The question arises how to make one of the right-handed neutrinos so light compared to the other two. One obvious possibility is to start with a flavor symmetry that predicts one of the masses to be zero. Small breaking of the symmetry will then generate a small but non-zero mass. This simple idea has been pursued in the context of keV dark matter 
\cite{Shaposhnikov:2006nn,Lindner:2010wr}, and will work for light sterile neutrinos as well. One possible symmetry that can be used is $L_e - L_\mu - L_\tau$ \cite{Petcov:1982ya,Petcov:2004rk}.  

\subsection{Minimal extended type I seesaw}
\label{sec:minimal}

The cases discussed in the last subsection have the feature that the fundamental seesaw Lagrangian already contains at least one particle at the desired eV scale. This is somewhat contradictory to the seesaw spirit, and one may ask whether it is also possible to generate eV-scale sterile neutrinos without the {\em a priori} presence of such states. 

Here we consider an interesting extension of the type I seesaw model,
in which three right-handed neutrinos together with {\it one} singlet $S$ are
introduced. A similar idea was used in Ref.~\cite{Chun:1995js}, with a sterile state of mass $\sim 10^{-3}$ eV introduced in order to explain the solar neutrino problem. Here we will show that there is a natural eV-scale sterile
neutrino in this scenario, without the need of inserting a small mass
term for $\nu_s$. The Lagrangian of this scenario is given by
\begin{eqnarray}\label{eq:L5}
-{\cal L}_m = \overline{\nu_L} M_D \nu_R  + \overline{S^c} M_S \nu_R
+ \frac{1}{2}\overline{\nu^c_R} M_R \nu_R +  {\rm h.c.}  ,
\end{eqnarray}
where $M_S$ is a $1 \times 3 $ matrix. The full $7 \times 7$
neutrino mass matrix in the basis $(\nu_L,\nu^c_R,S^c)$ reads
\begin{eqnarray}
M_\nu^{7 \times 7}  = \left(\begin{matrix}  0 & M_D & 0 \cr
M^T_D & M_R & M^T_S \cr 0 & M_S & 0
\end{matrix} \right)  .
\end{eqnarray}
In the case where $M_R \gg M_S > M_D $, one should first decouple
heavy right-handed neutrinos using the canonical seesaw formula, and
the effective neutrino mass matrix in the basis $(\nu_L,S^c)$ is given by
\begin{eqnarray}
M_\nu^{4\times 4} = - \left(\begin{matrix}  M_D  M^{-1}_R  M^T_D
& M_D  M^{-1}_R  M^T_S  \cr M_S  \left(M^{-1}_R\right)^T  M^T_D &
M_S M^{-1}_R M^T_S
\end{matrix} \right) .
\end{eqnarray}
Since $M_S$ is larger than $M_D$ by definition, one can apply the
seesaw formula once again and obtain
\begin{eqnarray}\label{eq:m5}
m_\nu \simeq M_D  M^{-1}_R  M^T_S \left(M_S  M^{-1}_R
M^T_S\right)^{-1} M_S  \left(M^{-1}_R\right)^T  M^T_D - M_D M^{-1}_R
M^T_D \, ,
\end{eqnarray}
for the active neutrinos, whereas there exists one sterile neutrino with mass
\begin{eqnarray}
m_s \simeq - M_S  M^{-1}_R M^T_S  \, .
\end{eqnarray}
Note that the right-hand-side of Eq.~\eqref{eq:m5} does not vanish since $M_S$ is a vector rather than a square matrix; if $M_S$ were a square matrix this would lead to an exact cancellation between the two terms of Eq.~\eqref{eq:m5}. 

The active-sterile neutrino mixing matrix takes a $4\times 4$ form, i.e.,
\begin{eqnarray}
V \simeq  \left(\begin{matrix} (1-\frac{1}{2}RR^\dagger)U & R \cr
-R^\dagger U & 1 -\frac{1}{2} R^\dagger R
\end{matrix} \right)   ,
\end{eqnarray}
where $R= M_D  M^{-1}_R  M^T_S \left( M_S  M^{-1}_R M^T_S\right)^{-1}$. Essentially, $V_{14}$ (i.e. $\sin\theta_{14}$) is suppressed by the ratio ${\cal O}(M_D/M_S)$.

As a naive numerical example, for $M_D \simeq 100~{\rm GeV}$, $M_S \simeq
500~{\rm GeV}$ and \mbox{$M_R \simeq 2\times 10^{14}~{\rm GeV}$}, one may
estimate that $m_\nu \simeq 0.05~{\rm eV}$, $m_s\simeq 1.3~{\rm eV}$
together with $R \simeq 0.2$. This is in very good agreement with the
fitted sterile neutrino parameters from Table~\ref{table:osc_params}.
Neutrino-less double beta decay is again allowed because not all of
the neutrinos are light. 

Note that one of the light neutrinos is massless since the rank of
${M_\nu}^{4\times 4}$ is three. This model is a minimal extension
of the type I seesaw in the sense that one needs at least three
heavy neutrinos to suppress the masses of both active and sterile
neutrinos. In other words, two heavy right-handed neutrinos give
rise to two massive active neutrinos, while the other one is
responsible for the mass of $\nu_s$. 

Unfortunately, if we embed this scenario into a grand unified theory
framework it cannot be gauge anomaly free since we have only one
generation of $S$. It is also not possible to accommodate two eV-scale
sterile neutrinos in this scenario, unless the number of heavy
neutrinos is increased. Apart from these shortcomings, the scenario
possesses the following  features:
\begin{itemize}

\item apart from the electroweak and seesaw scales, one does not
artificially insert small mass scales for
sterile neutrino masses. As in the canonical type I seesaw, one can 
take $M_S > M_D \sim {\cal O}(10^2 ~{\rm GeV})$, 
while $M_R$ can be chosen close to the
$B-L$ scale, not far from the grand unification scale; 

\item it is a minimal extension of the type I seesaw since three
heavy right-handed neutrinos can lead to at most three massive light
neutrinos (the ``seesaw-fair-play-rule''~\cite{Xing:2007uq}), 
out of which two are active and needed to account for the
solar and atmospheric neutrino mixing. It is more predictive owing to
the absence of one active neutrino mass, while it still accommodates 
all the experimental data; 

\item there exist heavy right-handed neutrinos that could be 
responsible for thermal leptogenesis. Note that, in the setup we
considered, right-handed neutrinos would preferably decay to the
sterile neutrino since their coupling to $S$ is larger than that to active
neutrinos. However, this drawback could be circumvented since
$S$ enters in the one-loop self-energy diagram of the decay of
right-handed neutrinos, which could compensate for this.

\end{itemize}

In this section we have exclusively used the type I seesaw and an
extension of it. 
We note that type II seesaw does not provide the possibility of light
sterile neutrinos, due to the absence of fermionic degrees of
freedom. Type III seesaw is formally analogous to type I, because 
the neutral components of the fermion triplets play the role of heavy
neutrinos. However, these components are not gauge singlets.

\section{conclusion}
\label{sec:summary}

Motivated by recently reported hints for sterile neutrinos from
particle physics, astroparticle physics and cosmology, we have 
studied the realization of light eV-scale sterile neutrinos in 
both model-independent and -dependent manners. 
The phenomenological consequences of light
sterile neutrinos were discussed and we pointed out that the
presence of such light sterile neutrinos would significantly change
the oscillation probabilities in short-baseline experiments as well as
the effective mass measured in neutrino-less double beta decay
experiments. It turns out that the sterile contributions to the 
effective mass could lead to a vanishing $\mee$ even if the three
active neutrinos obey an inverted ordering, which is clearly 
different from the predictions of the standard picture of three active neutrinos. Scenarios in which the
sterile states are heavier than the active ones are safest in what
regards cosmological mass limits; the other cases turn out to have
rather large contributions, and are even in danger of being ruled out by current limits on the effective mass. 

We then focused on the possibility of embedding sterile neutrinos into an effective theory with the $A_4$ flavor symmetry, and showed the corrections to the TBM pattern that result from this. We found that light sterile neutrinos can naturally be accommodated in the flavor $A_4$ symmetry, and their mixing with active neutrinos is correlated to the deviation from exact TBM. The departure from the initial mixing scheme is a general feature of such approaches; we only studied one example in
detail. Furthermore, we have shown that light sterile neutrinos can be realized in extensions of the seesaw model. Regardless of naturalness, the type I seesaw with three right-handed neutrinos could, in principle,
provide candidates for light sterile neutrinos. Besides the type I seesaw framework, we have presented a minimal extended type I seesaw model, in which, without the need of introducing tiny Yukawa couplings, the smallness of sterile neutrino masses is ascribed to the existence of heavy singlet neutrinos, whereas the mixing between active and sterile neutrinos could still be sizable. Note that all of the models we have discussed share the feature that the sterile neutrinos are heavier than the active ones. 

As mentioned before, we only concentrate on eV-scale sterile neutrinos in the current study. In general, a keV sterile neutrino, which may lead to rich phenomenological consequences in the early Universe and supernova explosions, could also be accommodated in the models discussed in this work. A detailed survey could be useful and will be elaborated on in future.

\begin{acknowledgments}
This work was supported by the ERC under the Starting Grant MANITOP and by the Deutsche Forschungsgemeinschaft in the Transregio 27 ``Neutrinos and beyond -- weakly interacting particles in physics, astrophysics and cosmology''. JB thanks M.~Holthausen, K.L.~McDonald and T.~Schwetz for useful discussions.
\end{acknowledgments}

\renewcommand{\theequation}{A-\arabic{equation}}
\setcounter{equation}{0}
\begin{appendix}

\section{Short-baseline neutrino oscillations: 3+2/2+3 vs.~1+3+1}

When neutrinos propagate in vacuum, the probability of finding a
neutrino of initial flavor $\alpha$ in the flavor state $\beta$ at
the baseline $L$ is given by
\begin{eqnarray}\label{eq:p}
P_{\alpha\beta} & = & \left| \langle \nu_\beta (L) |\nu_\alpha (0)
\rangle \right|^2 =  \left| \sum_i U_{\beta i} U^*_{\alpha i}
e^{-{\rm i} E_i L} \right|^2  \\
& = & \delta_{\alpha\beta} -4 \sum_{i>j}{\rm Re} \left( U^*_{\alpha
i}U_{\alpha j} U_{\beta i} U^*_{\beta j} \right)
\sin^2\left(\frac{\Delta_{ij}L}{4E}\right) + 2 \sum_{i>j}{\rm Im}
\left( U^*_{\alpha i}U_{\alpha j} U_{\beta i}  U^*_{\beta j} \right)
\sin\left(\frac{\Delta_{ij}L}{2E}\right) \, , \nonumber
\end{eqnarray}
where $\Delta_{ij} =m^2_i -m^2_j$. The current global-fit of neutrino
oscillation experiments indicates that (at $2\sigma$)
$\dms \equiv \Delta_{21}=(7.24\ldots 7.99) \times 10^{-5}~{\rm eV}^2$ and 
$\dma \equiv |\Delta_{31}|=(2.17\ldots 2.64) \times 10^{-3}~{\rm eV}^2$ and that
the corresponding mixing angles are $\sin^2\theta_{12}=0.28 \ldots
0.35 $, $\sin^2\theta_{23}=0.41\ldots 0.61$ and
$\sin^2\theta_{13}<0.031$~\cite{Schwetz:2011qt}, where we have
neglected small differences between normal and inverted ordering,
which are irrelevant for our discussion. The parameters
associated with sterile states are given in Table
\ref{table:osc_params}. 
The oscillation probability for anti-neutrinos can be simply obtained
from Eq.~\eqref{eq:p} by the replacement $U \to U^*$. 

It should be stressed that for a terrestrial based neutrino
oscillation experiment the baseline length is usually fixed, which
allows one to make reasonable expansions of the oscillation
probabilities. In a reactor anti-neutrino oscillation experiment, e.g.
Double Chooz or Daya~Bay, the detector is typically located at a baseline of around 1 km, while the average energy of
neutrinos is $\bar E \sim 4 ~{\rm MeV}$. 
One can thus estimate that $L/E \sim {\cal O}(10^3)~{\rm eV}^{-2}$,
indicating that the oscillation term containing $\Delta_{21}$ can be safely ignored. In addition, the
oscillation frequency related to the mass-squared difference
$\Delta_{4i}$ ($\Delta_{5i}$) is much smaller than the
baseline, and thus the $\Delta_{4i}$ ($\Delta_{5i}$) term
generates a fast oscillation. 
The anti-neutrino survival probability then approximates to
\begin{eqnarray}
P_{\overline{\nu}_e \rightarrow\overline{\nu}_e } & \simeq & 1 - 2
\left|U_{e4}\right|^2 - 4 \left|U_{e3}\right|^2
\sin^2\left(\frac{\Delta_{31}
L}{4E}\right) \nonumber \\
 &= & 1 - 2 \sin^2\theta_{14}-
\cos^4\theta_{14}\sin^2\left(2\theta_{13}\right)\sin^2\left(\frac{\Delta_{31}
L}{4E}\right) \, ,
\end{eqnarray}
for the case of one sterile neutrino (1+3 and 3+1), and
\begin{eqnarray}
P_{\overline{\nu}_e \rightarrow\overline{\nu}_e} & \simeq & 1 - 2
\left|U_{e4}\right|^2 -2 \left|U_{e5}\right|^2 - 4
\left|U_{e3}\right|^2 \sin^2\left(\frac{\Delta_{31} L}{4E}\right) \,
,
\end{eqnarray}
for the case of two sterile neutrinos (3+2, 2+3 and 1+3+1).  
It is clearly seen that the effect of the sterile neutrino(s) is
merely to reduce the total neutrino flux, which is needed to explain
the recently reported reactor anti-neutrino anomaly. 
Note that if in the Double~Chooz experiment the neutrino flux is
calibrated using the near detector instead of a Monte~Carlo
simulation, the total shift of the neutrino flux is cancelled
when comparing the number of events in the near and far detectors, 
thus having no effect on the measurement of $\theta_{13}$. 

In order to observe the oscillation effects induced by sterile 
neutrinos at a reactor neutrino experiment, a very short baseline, 
e.g., $L \sim {\cal O}(10~{\rm m})$, turns out to be very attractive. 
At such a short baseline length, the $\Delta_{31}$ contributions to
the oscillation probability can be neglected, and the survival probability reads
\begin{equation}
P_{\overline{\nu}_e \rightarrow\overline{\nu}_e} \simeq 1 -4
\left|U_{e4}\right|^2 \sin^2\left(\frac{\Delta_{41} L}{4E}\right) =
1 -\sin^2\left(2\theta_{14}\right) \sin^2\left(\frac{\Delta_{41}
L}{4E}\right) \, ,
\end{equation}
for the 3+1/1+3 case, and
\begin{equation}
P_{\overline{\nu}_e \rightarrow\overline{\nu}_e} \simeq 1 -4
\left|U_{e4}\right|^2 \sin^2\left(\frac{\Delta_{41} L}{4E}\right)-4
\left|U_{e5}\right|^2 \sin^2\left(\frac{\Delta_{51} L}{4E}\right) \,
,
\end{equation}
for the 2+3, 3+2 or 1+3+1 cases. Note that the oscillation term
proportional to $\sin^2\left(\frac{\Delta_{54} L}{4E}\right)$ is 
suppressed by a factor $\left|U_{e4}U_{e5}\right|^2$, and
can hence be neglected. The practical realization of a detector so
close to a reactor is not straightforward, and will not be discussed
here (see Ref.~\cite{Vergados:2011na} for a recent proposal). So far there is no possibility to distinguish 2+3/3+2 from 1+3+1
scenarios. 

The LSND and MiniBooNE experiments report an excess of electron anti-neutrino events in the oscillation channel
$\overline{\nu}_\mu \rightarrow \overline{\nu}_e$, with a baseline and
energy setup $L/E \sim {\cal O}(1)~{\rm eV}^{-2}$. In analogy to the
very short baseline reactor neutrino oscillations, 
the oscillation term related to $\Delta_{31}$ can be ignored, and the
transition 
probability reads
\begin{eqnarray}
P_{\overline{\nu}_\mu\rightarrow\overline{\nu}_e} \simeq 4
|U_{e4}|^2|U_{\mu 4}|^2 \sin^2\left(\frac{\Delta_{41} L}{4E}\right)\, ,
\end{eqnarray}
for the 3+1/1+3 case, and
\begin{eqnarray}\label{eq:P5}
P_{\overline{\nu}_\mu\rightarrow\overline{\nu}_e} & \simeq & 4
|U_{e4}|^2|U_{\mu 4}|^2 \sin^2\left(\frac{\Delta_{41} L}{4E}\right)
+ 4 |U_{e5}|^2|U_{\mu 5}|^2 \sin^2\left(\frac{\Delta_{51}
L}{4E}\right) \nonumber \\
& + &  8 |U_{e4} U_{\mu 4} U_{e5} U_{\mu 5}|
\sin\left(\frac{\Delta_{41} L}{4E}\right)\sin\left(\frac{\Delta_{51}
L}{4E}\right)\cos\left(\frac{\Delta_{54} L}{4E} + \delta\right)\, ,
\end{eqnarray}
for the five neutrino case with $\delta \equiv {\rm arg}\left(U^*_{e4}
U_{\mu 4} U_{e5} U^*_{\mu 5} \right) \simeq \delta_{14} - \delta_{24}
- \delta_{15}$, where we have given the phase in the explicit parameterization of
Eq.~(\ref{eq:U2st}). The additional CP violating
phase in Eq.~\eqref{eq:P5} leads to a CP asymmetry in the transition
probability between the neutrino and anti-neutrino modes, which is of
particular interest in view of the null oscillation results from the
neutrino running of MiniBooNE. In addition, the last term of
Eq.~\eqref{eq:P5} indicates a distinctive difference between the
3+2/2+3 and 1+3+1 cases due to the presence of $\Delta_{54}$.  
Explicitly, in the 3+2/2+3 cases one has 
$\Delta_{54}=|\Delta_{51}| - |\Delta_{41}|$ whereas in the 1+3+1 
case, $\Delta_{54} \simeq |\Delta_{51}| + |\Delta_{41}|$, which
changes the transition probability regardless of the choice of CP violating phases.

\begin{figure}[t]\vspace{-0.2cm}
\includegraphics[width=165mm]{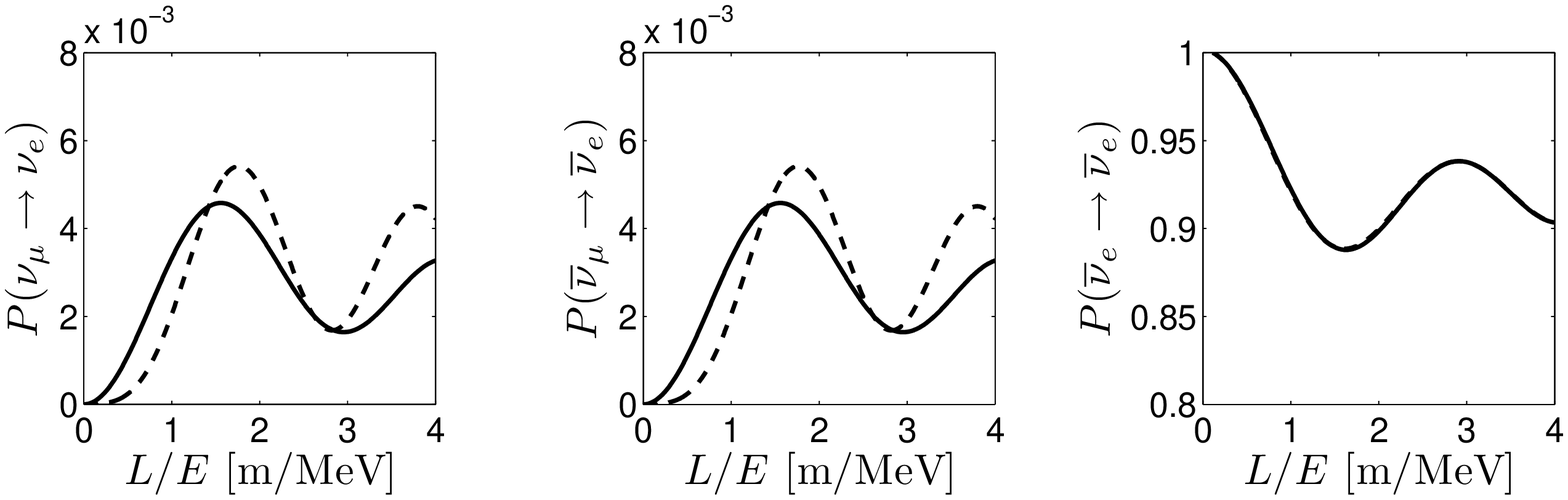}\vspace{-0.8cm}
\includegraphics[width=165mm]{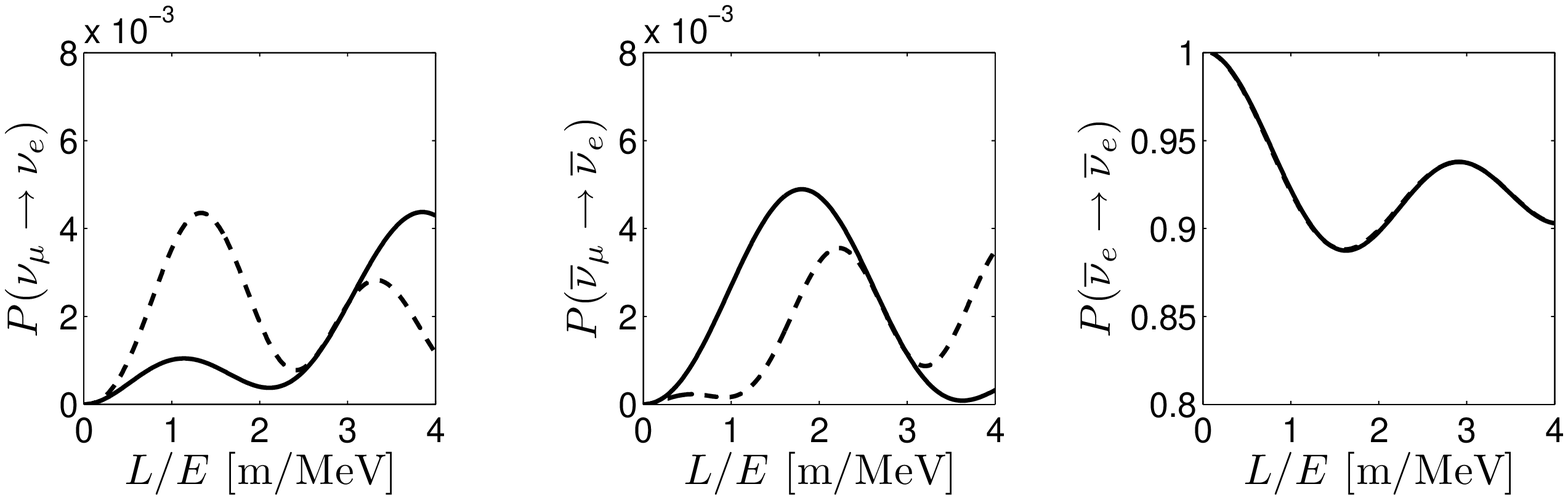}\vspace{-0.5cm}
\caption{\label{fig:fig1} The oscillation probabilities with respect
to the quantity $L/E$ in the 3+2/2+3 cases (solid lines) and 1+3+1
case (dashed lines). The mixing parameters are the best-fit values from TABLE II
of Ref.~\cite{Schwetz:2011qt} and TABLE II of Ref.~\cite{Kopp:2011qd}. In the upper panel, we set all 
CP violating phases to be zero, whereas in the lower panel $\delta =
\frac{\pi}{2}$ is assumed.}
\end{figure}

We illustrate the oscillation probabilities with respect to the ratio
$L/E$ in Fig.~\ref{fig:fig1} for the 3+2/2+3 and 1+3+1 cases. 
It is clear that the transition probabilities are quite different in
the hierarchy schemes, whereas the survival probability
is not sensitive to the sterile neutrino hierarchies since the
$\Delta_{54}$ contributions are suppressed. Furthermore, if
non-vanishing CP violating phases are included there are visible
differences between neutrino and anti-neutrino flavor transitions,
implying that a detector at very short distance would be an ideal
place to search for the CP violation related to sterile neutrinos.

\end{appendix}

\bibliography{bib}

\end{document}